\documentclass[fleqn,usenatbib]{mnras}

\usepackage{newtxtext,newtxmath}

\usepackage[T1]{fontenc}

\DeclareRobustCommand{\VAN}[3]{#2}
\let\VANthebibliography\thebibliography
\def\thebibliography{\DeclareRobustCommand{\VAN}[3]{##3}\VANthebibliography}

\usepackage{amsmath}
\usepackage{multirow}
\usepackage{subcaption}
\usepackage{caption}
\usepackage{threeparttable}
\usepackage{graphicx}	
\usepackage{amsmath}	

\newcommand\mrm{\mathrm}
\newcommand\mbh{M_{\mrm{BH}}}

\newcommand\rsr{R_{\star}}
\newcommand\msr{M_{\star}}
\newcommand\rp{R_{\mrm{p}}}
\newcommand\re{R_{\mrm{e}}}

\newcommand\rs{R_{\mrm{S}}}
\newcommand\rsun{R_{\odot}}
\newcommand\msun{M_{\odot}}
\newcommand\dmfb{\dot{M}_{\mrm{fb}}}
\newcommand\dntdeo{\dot{N}_{\mrm{TDE,0}}}
\newcommand\dntde{\dot{N}_{\mrm{TDE}}}
\newcommand\ledd{L_{\mrm{Edd}}}

\newcommand\tvis{T_{\mrm{viscous}}}

\newcommand\ugriaz{u,\ g,\ r,\ i \mrm{\ and\ } z}

\newcommand\mgaltdf{m_{\mrm{gal+TDF}}}

\title[Prospects of Finding TDEs with WFST]{The Prospects of Finding Tidal Disruption Events with 2.5-Metre Wide-Field Survey Telescope (WFST) Based on Mock Observations}

\author[Lin, Jiang \& Kong]{
Zheyu Lin,$^{1,2}$\thanks{E-mail: linzheyu@mail.ustc.edu.cn}
Ning Jiang,$^{1,2,3}$\thanks{E-mail: jnac@ustc.edu.cn}
and Xu Kong$^{1,2,3}$\thanks{E-mail: xkong@ustc.edu.cn}
\\
$^{1}$Department of Astronomy, University of Science and Technology of China, 
Hefei, 230026, China\\
$^{2}$School of Astronomy and Space Sciences, 
University of Science and Technology of China, Hefei, 230026, China\\
$^{3}$Frontiers Science Centre for Planetary Exploration and Emerging Technologies, University of Science and Technology of China, Hefei, Anhui, 230026, China
}

\date{Accepted 2022 March 31. Received 2022 March 17; in original form 2022 January 11}

\pubyear{2022}

\begin{document}
\label{firstpage}
\pagerange{\pageref{firstpage}--\pageref{lastpage}}
\maketitle

\begin{abstract}
Optical time-domain survey has been the dominant means of hunting for rare tidal disruption events (TDEs) in the past decade and remarkably advanced the TDE study. Particularly, the Zwicky Transient Facility (ZTF) has opened the era of population studies and the upcoming Large Synoptic Survey Telescope (LSST) at the Vera Rubin Observatory (VRO) is believed to further revolutionize the field soon. Here we present the prospects of finding TDEs with another powerful survey to be performed by 2.5-metre Wide-Field Survey Telescope (WFST). The WFST, located in western China, will be the most advanced facility dedicated to optical time-domain surveys in the northern hemisphere once commissioning. We choose to assess its TDE detectability on the basis of mock observations, which is hitherto closest to reality by taking into consideration of site conditions, telescope parameters, survey strategy and transient searching pipeline. Our mock observations on 440 deg$^2$ field (CosmoDC2 catalogue) show that $29\pm6$ TDEs can be robustly found per year if observed at $u, g, r, i$ bands with 30-second exposure every 10 days, in which a discovery is defined as $\geq$10 epochal detections in at least two filters. If the WFST survey is fully optimized for discovering TDE, we would expect to identify $392\pm74$ of TDEs every year, with the redshift up to $z\sim0.8$, which poses a huge challenge to follow-up resources.
\end{abstract}

\begin{keywords}
telescopes -- transients: tidal disruption events
\end{keywords}



\section{Introduction} \label{sec:intro}

\subsection{Tidal Disruption Events (TDEs)}

An unlucky star in the core of a galaxy which wanders too close to the supermassive black hole (SMBH) can be torn apart if the tidal force of BH exceeds the self-gravity of the star \citep{ 1988Natur.333..523R,1989ApJ...346L..13E,1989IAUS..136..543P}. After the disruption, about half of the stellar debris escapes from the BH, while the remaining half is accreted by the BH, producing a luminous flare of electromagnetic emission lasting for months to years. At the peak luminosity, the flare can light up the central region of the galaxy, and even outshine the whole host galaxy. This phenomenon is called as a tidal disruption event (TDE) and has aroused extensive interests because of its unique scientific values. First of all, TDE has provided us an excellent opportunity to investigate the SMBHs in normal galaxies \citep[e.g.,][]{2019ApJ...872..151M,2019Sci...363..531P}, which is otherwise extremely difficult to probe, particularly for dwarf and faraway galaxies. TDEs could even unveil dormant intermediate-mass BHs (IMBHs) \citep{2018NatAs...2..656L} and SMBH binaries \citep{2014ApJ...786..103L,2020NatCo..11.5876S}. Moreover, TDEs can also serve as an ideal laboratory for exploring the accretion physics of SMBHs, i.e., those unsettled issues in active galactic nuclei (AGNs), by observing the whole life cycle of BH activity. Recent progresses in infrared echoes suggest that TDEs have also offered an effective means to study the sub-parsec environments of quiescent SMBHs \citep[e.g.][]{2016ApJ...828L..14J,2016ApJ...829...19V,2021ApJ...911...31J}. 

However, the observational discoveries of TDEs are quite challenging because of its rather low event rate, which is at orders of $10^{-4}-10^{-5}$ yr$^{-1}$ for most galaxies \citep[e.g.,][]{2002AJ....124.1308D,2008ApJ...676..944G,2014ApJ...792...53V,2018ApJ...852...72V}. Motivated by the theoretical predication of peak spectral energy distributions (SEDs) at soft X-ray or extreme ultraviolet (EUV), TDEs were first identified out as soft X-ray transients in the galactic nuclei with the archival ROSAT data in late 1990s \citep[e.g.][]{Bade1996,1999A&A...343..775K}. Subsequently, other X-ray instruments, i.e., XMM-Newton, Chandra and Swift, also joined in the journey of discovery one after another \citep[e.g.][]{2007A&A...462L..49E,2010ApJ...722.1035M,2011Sci...333..203B,2020SSRv..216...85S}, accompanied with a few candidates found in UV bands \citep[e.g.][]{2006ApJ...653L..25G}. Nevertheless, they were all found serendipitously from archival data and thus have scarce synergetic information in other wavelength regimes, which seriously hampered the progress of the field. Until very recently, the SRG/eROSITA all sky survey has shown the power of discovering X-ray TDEs in bulk but with only half-year cadence of light curves \citep{2021MNRAS.508.3820S}.

The number growth of TDEs in the past decade is mainly owing to a variety of wide-field time-domain surveys in optical bands (see recent review of \citealt{2020SSRv..216..124V,Gezari2021}), such as Pan-STARRS (\citealt{Gezari2012}), PTF (\citealt{Arcavi2014}), and ASAS-SN (\citealt{Holoien2016}), even though the origin of optical emission is still under debate (see review of \citealt{Roth2020} and recent models of \citealt{Lu2020,Liu2021}) . More importantly, these events are noticed in real time, making timely multi-wavelength follow-up observations realistic. Particularly with the commission of the Zwicky Transient Facility \citep[ZTF,][]{2019PASP..131a8002B}, the cumulative discovery rate of TDEs has increased from $\lesssim$2/yr to $>$10/yr, opening the era of statistical research \citep{2021ApJ...908....4V}. The dominance of uncovering TDEs in optical bands will be continued, as the Large Survey of Space and Time (LSST) at the Vera Rubin Observatory (VRO) is scheduled to start survey at 2023 (\citealt{2019ApJ...873..111I}). The wide-field and high-cadence survey with 8.4-metre primary mirror telescope ensures VRO to be a milestone for time-domain survey (\citealt{2009arXiv0912.0201L}). 
Works from different groups have all predicted that at least thousands of TDEs ($\sim$1,600$-$8,000) per year can be discovered by VRO/LSST (\citealt{2009MNRAS.400.2070S,2011ApJ...741...73V,2015ApJ...814..141M,2019MNRAS.488.4042T,2020ApJ...890...73B,2021ApJ...910...93R}). 

The unprecedented VRO/LSST will mainly scan the southern celestial sphere while ZTF covers the northern celestial sphere, yet with a much shallower depth. To complement VRO in sky coverage and ZTF in imaging depth, a new time-domain survey is planned with a
2.5-metre Wide-Field Survey Telescope (WFST) in China, which is competitive in discovering various transients. In the following, we will introduce the WFST in details first and then try to estimate its detection rates of TDEs based on mock time-series images.

\subsection{Wide-Field Survey Telescope (WFST)}

The WFST is being constructed by the University of Science and Technology of China (USTC) and the Purple Mountain Observatory (PMO), and will begin to survey the northern sky at late 2022. This facility is characterized by a 2.5-metre primary mirror and a primary-focus camera with field of view (FOV) of 7 square degrees filled with a 
9 pieces of $9\textrm{K}\times9\textrm{K}$ mosaic CCD detector, yielding out an effective FOV of 6.55 square degrees in CCD. The key scientific goals of WFST include: 
1) survey the northern sky with the highest sensitivity to explore the variable universe and catch up the time-domain events;
2) find and track one million solar system objects for a panchromatic view of the
solar system and understand its kinematic evolution; discover planets or their moons in the Kuiper Belt and beyond; 
3) provide high-precision astrometric and photometric catalogues of objects down to $m_r < 25$, allowing precise mapping of the structures of the Milky Way and the nearby universe, which are either science breakthrough such as gravitational events or unknown in the current framework of astrophysics.

During the design study of the WFST, an optical design based on a primary-focus system has been developed. This design features a wide FOV, low obscuration, high efficiency, wide band coverage and high image quality. 80\% of the energy of a point source falls within 0.40 arcsec across the full 3-degree field and for all bands. During the design study, special efforts
have been paid to increase the system transmission at the $u$ band. This has a great impact on the design choice for the atmospheric dispersion compensator (ADC). To maintain the image quality in real time, an active optics system based on
laser tracker measurements and curvature wavefront sensors is proposed. The design of the WFST can be found in \citet{2016SPIE10154E..2AL}, we briefly introduce the optical design, active optics, and the focal-plane instrument in the following passages.

\begin{figure*}
    \centering
    \includegraphics[width=0.9\textwidth]{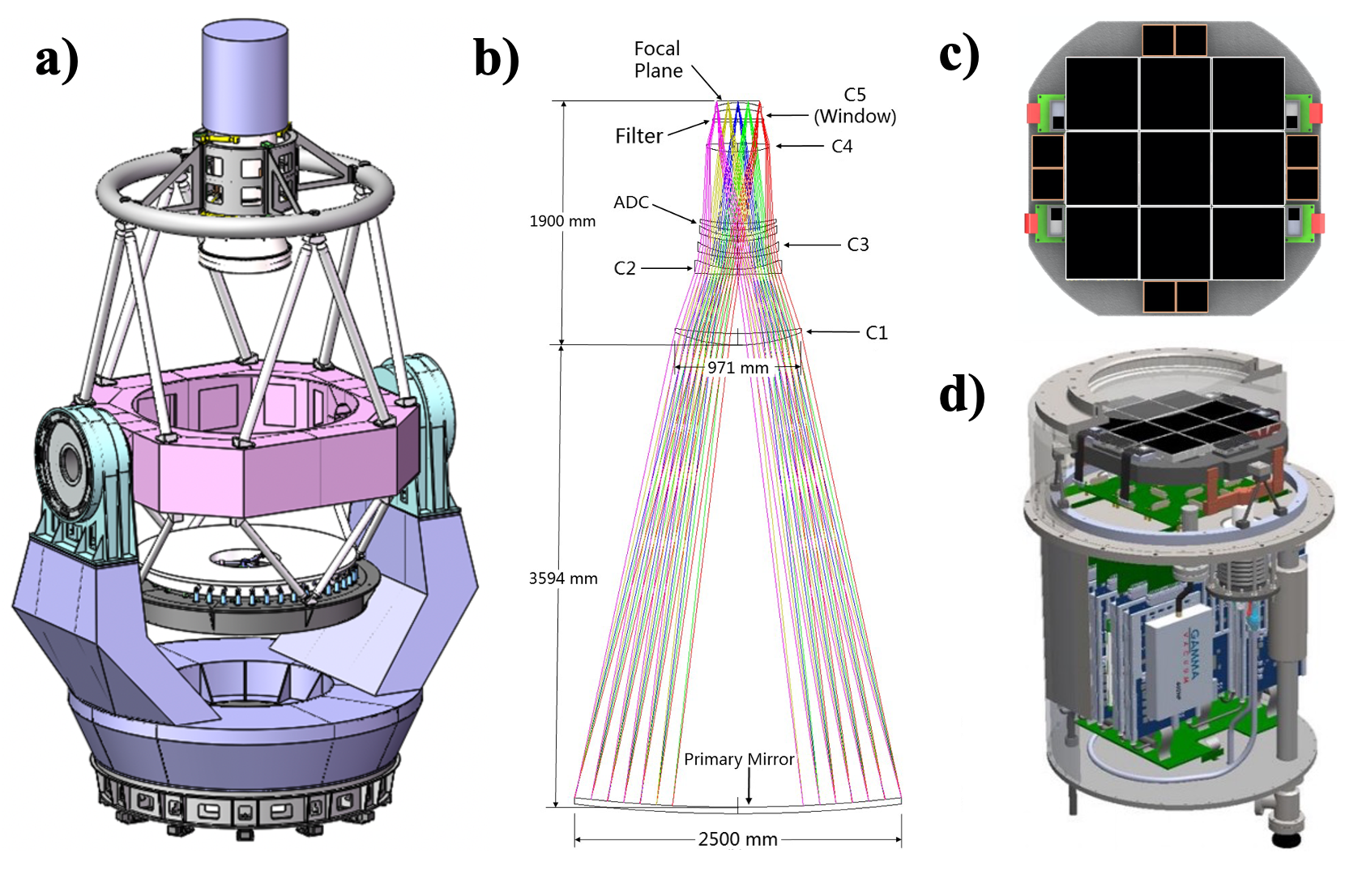}
    \caption{Panel a): 3D model of the WFST; Panel b): Optical layout of the WFST; Panel c): The WFST camera provides a 0.765 Gigapixel flat focal plane array, tiled by 9 pieces of $9\textrm{K}\times9\textrm{K}$ CCD science sensors with 10 $\mu$m pixels (white borders). Eight additional $4\textrm{K}\times4\textrm{K}$ chips are used for curvature wavefront sensors (orange borders), and four additional chips are used as guiding sensors (green); Panel d): A vacuum cryogenic dewar that houses the CCD chips, the RAFT structures, and the readout electronics. It provides the required working temperature of $-100^{\circ}$C for the CCDs.}
    \label{fig:wfst}
\end{figure*}

A 3D model of the WFST is shown in Figure \ref{fig:wfst}a. The outlook of the telescope is dictated by its primary-focus optical layout. The focal-plane instrument and the corrector lenses are mounted as an integrated unit located near the focus of the primary mirror, which is called the primary-focus assembly (PFA). 
The PFA sits on the top end of the telescope tube via a hexapod, which provide both support and position alignment to the PFA. 
The primary mirror and mirror cell are located on the other end of the tube. The telescope tube itself is assembled on an altazimuth mount. The top-level telescope specifications are listed in Table \ref{tab:wfst}.

The optical layout of the WFST is shown in Figure \ref{fig:wfst}b. The optical system consists of 
the primary mirror, 5 corrector lenses, an ADC and six filters. 
The primary mirror is an $F$/2 aspheric surface with a diameter of 2.5 m. After the corrector lenses, the final system focal ratio becomes $F$/2.48. 
The primary surface is a hyperbolic concave surface with high-order aspheric terms. 
The first corrector lens is the largest lens with a diameter of 970 mm. The lens is made of fused silica and has a meniscus shape with a central thickness of 90 mm. 
The second corrector lens is a negative lens whose concave surface is also a high-order aspheric surface. The third and fourth lenses are all spherical lenses. To control the chromatic aberrations, the third lens is made of Schott’s N-BK7 glass. 
A pair of lensm (lens with a wedge) ADC is inserted between the third and fourth lenses to compensate for the atmospheric dispersion. 
The fifth lens is a field corrector, which corrects field curvature and reduces distortions. It also serves as the barrier window for the detector dewar. 

The WFST PFA is comprised of the mosaic camera, a rolling shutter, filters and filter switch mechanism, and an image rotator. 
The WFST camera provides a 0.765 Gigapixel flat focal plane array, tiled by 9 pieces of $9\textrm{K}\times9\textrm{K}$ CCD science sensors with 10 $\mu$m pixels (CCD290-99 from e2V). This pixel count is a direct consequence of sampling the 6.55 deg$^2$ FOV (0.325 m diameter) with $0.333\times0.333$ arcsec$^2$ pixels. 
Eight additional $4\textrm{K}\times4\textrm{K}$ chips are used for curvature wavefront sensors (surrounded by orange borders in Figure \ref{fig:wfst}c) and four additional chips are used as guiding sensors (shown as green in Figure \ref{fig:wfst}c). 
The CCD chips, the RAFT structures, and the readout electronics are housed in a vacuum cryogenic dewar, which provides the required working temperature of $-100^{\circ}$C for the CCDs (shown as green in Figure \ref{fig:wfst}d).
The entrance window to the cryostat is the fifth of the five refractive lenses in the camera. 
The telescope operates at six wavelength bands ($u,\ g,\ r,\ i,\ z$ and $w$), spanning from 320 to 925 nm. 
An image rotator provides rotation mechanism for the dewar and shutter with respect to the outer frame of the camera. 
The camera weighs about 460 kg, and its axial position tolerance with respect to the corrector lenses is only 2 $\mu$m. 
To reduce the burden on the hexapod that supports the PFA, the camera itself is equipped with a focusing motor, which is able to provide an extra axial adjustment up to $\pm 2$ mm.

\begin{table*}
\caption{Top-level specifications for WFST\label{tab:wfst}}
\centering
\begin{tabular}{c|c}
\hline\hline
\textbf{Item} & \textbf{Specifications}\\ \hline
Optical Configuration & Primary-focus with corrector lenses \\ \hline
Aperture & 2.5 m diameter \\ \hline
Focal Length & 6.2 m \\ \hline
Field of View & 3 deg diameter \\ \hline
Etendue & 29.3 m$^2$ deg$^2$ \\ \hline
Wavelength & 320-925 nm ($u,\ g,\ r,\ i,\ z,\ w$) \\ \hline
Image Quality & Diameter $\leqslant$ 0.4 arcsec (80\% energy encircled) \\ \hline
Plate Scale & 30 mm/arcsec \\ \hline
Pixel Size & 10 $\mu$m $\times$ 10$\mu$m \\ \hline
Number of pixels & 0.765 Giga \\ \hline
\multirow{2}{*}{Survey Depth} & $m_{u}=22.31,\ m_{g}=22.93,\ m_{r}=22.77,\ m_{i}=22.05,$\\ 
& $m_{z}=21.02,\ m_{w}=22.96$ for 30s exposures \\ \hline
\end{tabular}
\end{table*}

The site of WFST is on the top of Saishiteng Mountain near the town of Lenghu, Qinghai Province. Its geographical coordinates are 38.6068° N, 93.8961° E, and it has an elevation of 4,200 m.
The three-year monitoring at a local summit on Saishiteng Mountain has proven it being a world-class astronomical observing site, which have 70\% clear nights, median seeing of 0.75 arcseconds and median night sky background of 22.0 mag/arcsec$^2$ \citep{2021Natur.596..353D}.
Moreover, the Lenghu site in the Eastern Hemisphere, has bridged the huge gap between Mauna Kea, Atacama and the Canary Islands, which will constitute a global network of large observatories ready for great scientific discoveries in the golden time-domain era. 

Thanks to its elaborate design and the top-level site condition, WFST is expected to be an excellent hunter for various optical transients, especially the rare ones which can be only captured by advanced time-domain surveys. According to \citet{2018AcASn..59...22S}, the 5-$\sigma$ limiting magnitude for a 30-second exposure in each band is: $m_{u}=22.31,\ m_{g}=22.93,\ m_{r}=22.77,\ m_{i}=22.05,\ m_{z}=21.02,\ m_{w}=22.96$. 

In this work, we choose to focus on the detectability of TDEs quantitatively with mock observations by taking into account of different factors. The methods and procedure involved are introduced in Section \ref{sec:conmocki} and \ref{sec:mocko}. Then we will show the main results from mock observations in Section \ref{sec:results} and end with a brief conclusion in Section \ref{sec:disc}.

\section{Method} \label{sec:conmocki}

In this section, we introduce the parameterized models used to generate light curves and host galaxies of TDEs.  







\subsection{Galaxy Catalogue}
\label{galcatalogue}

The demand of host galaxy information for mock observation is three-folds. First, in reality, a TDE candidate is identified only when it is associated with the centre of a galaxy. Second, the emission of tidal disruption flare (TDF) will be unavoidably diluted by its host galaxy, in the form of background contamination and dust extinction. Third, the host galaxy could provide additional parameters required by the {\tt MOSFiT} TDE model (see Section \ref{sec:mosfit}). 



We finally choose CosmoDC2, a synthetic galaxy catalogue that contains $\sim$2.26 billion galaxies in a sky area of 440 deg$^2$ \citep{2019ApJS..245...26K}. 
The galaxies in this field match with the expected number densities from contemporary VRO/LSST surveys to $m_{r}=28$ and $z=3$. Each galaxy comprises an exponential disk component and a bulge component with Sérsic index $n=4$. In addition, the redshift $z$ and central black hole mass $\mbh$ are both given for each galaxy, and the rest-frame spectral energy distributions (SEDs) from 100 to 2,000 nm for each component are also provided, enabling the calculation for fluxes and AB magnitudes in WFST bands. Moreover, the morphological parameters needed to reconstruct the disk and bulge images (e.g., half-light radius $\re$, ellipticity) are all included by the catalogue. 



The galaxy number of the whole catalogue is however tremendous. 
We have first abandoned those which are obviously beyond the parameter spaces considered in this work. 
The detailed steps are presented as below.

\begin{enumerate}
    \item Exclude galaxies at $z>1.5$, as TDEs at such high redshift are too faint to be observed. This cut will get rid of a large fraction of galaxies since most galaxies in the catalogue are located at $1.5<z<3$ actually.
    \item Exclude galaxies with $\mbh<10^5\,\msun$. Theoretically, TDEs can occur in these galaxies, e.g., tidal disruption of white dwarfs by intermediate-mass black holes (see review by \citealt{Maguire2020}). However, their physical details, and thus their anticipated optical light curves remain rather unclear. 
    \item Exclude galaxies with $\mbh>M_{\mrm{Hills}}$. The Hills mass $M_{\mrm{Hills}}$ is the maximum mass for a Schwarzschild BH, whose event horizon keeps inside of the tidal radius.
    It is defined as
    \begin{equation}
    M_{\mrm{Hills}}=9.0\times10^7\,\msun\left(\frac{\msr}{\msun}
    \right)^{-1/2}\left(\frac{\rsr}{\rsun}\right)^{3/2}
    \end{equation}
    \citep{1975Natur.254..295H,1992MNRAS.259..209B,2016NatAs...1E...2L}. 
    We have applied a Kroupa IMF, whose minimum and maximum stellar masses are 0.08 $\msun$ and 1 $\msun$, respectively, and used the relation $\rsr\propto\msr^{0.8}$ on lower main sequence \citep{2016MNRAS.455..859S} to get $M_{\mrm{Hills}}\propto\msr^{0.7}$. As a result, galaxies with $M_{\mrm{BH}}\gtrsim9.0\times10^7\,\msun$ have been excluded since $M_{\mrm{BH}}$ exceeds the maximum Hills mass in this mass range.
\end{enumerate}

The three criteria above have readily rejected most galaxies, with
the remaining number decreases down to about 52 million.




\subsection{Selecting Galaxies hosting TDEs} \label{sec:select}

The galaxy number after the primary cut above is still very large and 
will be extremely time-consuming if we model all of them.
In fact, the TDE rate for a given galaxy is very low, meaning that TDEs will occur only in a very small fraction of these galaxies within one year. 
Therefore, we choose to first pick out the galaxies with real TDE 
occurrence in the mock observations and neglect others.
The adopted TDE rate is the one given by \citet{2016MNRAS.455..859S} with the modification of cosmological effect, that is
\begin{eqnarray}
\dntde(z) & = & \dntdeo\frac{\mrm{d}t_0}{\mrm{d}t}\nonumber\\
& = & 10^{-4.19}\frac{1}{1+z}\left(\frac{\mbh}{10^8\msun}\right)^{-0.223}\,\mrm{yr}^{-1}\label{eqn:dntde},
\end{eqnarray}
where $t_0$ and $t$ are the time in the frame of the source and the observer, respectively.

With knowing of the TDE occurrence rate of a specific galaxy assigned by Equation~\ref{eqn:dntde}, we then determine whether or not there happens 
a TDE within one year. Only if the galaxy has been picked out as a TDE host
in the simulation, we take it for further consideration. For simplicity, we assume one galaxy can undergo at most one TDE per year. 

We have performed the selection process for 100 times in order to minimize 
the random errors induced in the selection process. The averaged number obtained is 
5,192$\,\pm\,$62, that is an event rate of $\sim0.01\%$ for these galaxies. 
The reduced number up to this step becomes basically acceptable.

\subsection{TDF light curves generated by MOSFiT} \label{sec:mosfit}

The classical theory has predicted a $t^{-5/3}$ declination in the light curves of the tidal disruption flare (TDF),
which is purely determined by the mass fall-back rate of stellar debris after disruption. Observationally, optical surveys have yielded out dozens 
of TDEs with well-sampled light curves so far (\citealt{2020SSRv..216..124V,Gezari2021}). 
They show somewhat consistency but also discrepancy 
with the theoretical prediction.
Therefore, it is improper to create optical light curves crudely from theory 
and some empirical models have been developed to fit the 
observational data better.
The {\tt MOSFiT}, one of such kind of models, is exactly designed to help bridge 
the gap between observations and theories for different types of transients (\citealt{2018ApJS..236....6G}). 
In addition to fitting the light curves of known TDEs, {\tt MOSFiT} also yields statistically consistent predictions for other characteristics, for instance, 
the $\mbh$ (\citealt{2019ApJ...872..151M}), proving that it is a reliable
model.

Specifically, 11 parameters have been used to generate light curves by {\tt MOSFiT}. We introduce these parameters and their settings as follows:
\begin{enumerate}
    \item Redshift $z$.
    \item Black hole mass $\mbh$.\vspace{10pt}\\
    For $z$ and $\mbh$, we apply the values directly from the galaxy catalogue we use.\\
    \item Hydrogen column density $N_{\mrm{H}}$.\\
    {\tt MOSFiT} uses a ratio $N_{\mrm{H}}=1.8\times10^{21}A_V$ to convert $A_V$ into $N_{\mrm{H}}$. To calculate $A_V$ at the galaxy centres, we refer to the calculation method of \citet{2021ApJ...910...93R}. First, we obtain $A_{\rm H\alpha}$ for each galaxy. Galaxies that have specific star-formation rate sSFR $>10^{-11.3}$ yr$^{-1}$ are classified as star-forming galaxies, while the rest are labelled as quiescent galaxies. For star-forming galaxies, $A_{\rm H\alpha}$ is drawn from a Gaussian distribution with a floor at zero, where the median is described by an equation that only related to the stellar mass of the host galaxy,
    \begin{equation}
    \begin{aligned}
        A_{\rm H\alpha,median} &=0.91+0.77x+0.11x^2-0.09x^3\rm{, where}\\
        x&\equiv\rm{log}_{10}\left(\frac{\rm{host}\  \it{M_*}}{10^{10}\,\msun}\right),
    \end{aligned}
    \end{equation}
    and the standard deviation is 0.28 mag. This relationship was only calibrated for star-forming galaxies with total stellar mass between $10^{8.5}$ and $10^{11.5}\,\msun$ by \citet{2010MNRAS.409..421G}, therefore we use the edge values of the equation for star-forming galaxies with $M_*<10^{8.5}\,\msun$ and $M_*>10^{11.5}\,\msun$. For quiescent galaxies, we draw $A_V$ from a Gaussian distribution whose median and standard deviation are 0.2 and 0.06 mags, respectively, with a floor at zero. This choice of median is based on the results of \citet{2015A&A...581A.103G}. We then adopt the \citet{2000ApJ...533..682C} law and $R_V=4.2$ to convert all $A_{\rm H\alpha}$ into $A_V$. Finally, the host galaxy and the Galactic dust extinctions are applied to each event according to the model in \citet{1994ApJ...422..158O}.\\
    We realize that CosmoDC2 has provided dust extinction parameters $A_V$ and $R_V$. However, both parameters are found too low to describe TDEs that take place in galaxy centres. Therefore, we choose the 
    method of \citet{2021ApJ...910...93R} to estimate the dust extinction.
    
    \item The reprocessing layer can help explain the optical/UV emission of TDFs. In this model, it is assumed as a simple blackbody photosphere, and its radius has a power-law dependence on luminosity, defined as 
\begin{equation}
    R_{\mrm{phot}}=R_{\mrm{ph0}}\,a_{\mrm{p}}(L/\ledd)^l,
\end{equation}
     where $a_{\mrm{p}}$ is the semimajor axis of the accreting mass at peak $\dmfb$, and $\ledd\equiv4\pi G\mbh c/\kappa$ is the Eddington luminosity. This equation contains two free parameters: the power-law exponent $l$ and radius normalization $R_{\mrm{ph0}}$.
     \item Variance parameter $\sigma$.
     \item Explosion time $t_{\mrm{exp}}$, when the TDE starts.
     \item Viscous time $T_{\mrm{viscous}}$, defined as
\begin{equation}
\dot{M_\mrm{d}}(t)=\dot{M_\mrm{fb}}(t)-M_\mrm{d}(t)/T_{\mrm{viscous}},
\end{equation}
     where $M_\mrm{d}$ is the mass that remains suspended outside of the black hole’s horizon for roughly a viscous time, and $\dot{M}_{\mrm{d}}$ is the accretion rate onto the black hole from the forming disk.\vspace{10pt}\\
      For $l$, $R_{\mrm{ph0}}$, $\sigma$ and $T_{\mrm{viscous}}$, we refer to the fitting results from the 14 TDEs in \citet{2019ApJ...872..151M}, and set $l=1.5,$ log$_{10}(R_{\mrm{ph0}})=0.8$, $\ \sigma=0.1,\ \tvis=0.001-0.5$. 
    We have noticed that the light curve production needs at least one free parameter and $\tvis$ has been chosen by us. Actually, the light curves with different $\tvis$ values show extremely weak discrepancy in our experiments.
    Also, it should be emphasized that though the settings of $l=1.5,$ log$_{10}(R_{\mrm{ph0}})=0.8$ are the default settings of the {\tt MOSFiT} TDE model, and representative for known TDEs (\citealt{2019ApJ...872..151M}), a broader range of values 
    is possible and might affect our results as the light curve modelling is
    sensitive to $l$ and $R_{\mrm{ph0}}$. However, it is beyond the scope of this work.\\
    \item Radiative efficiency $\epsilon$.\\
    We set $\epsilon=0.1$. We note that the efficiency $\epsilon$ varies from 0.006 to 0.2 in the real fitting (\citealt{2019ApJ...872..151M}). However, we still simply adopt the most conventional value 0.1 given our poor knowledge of the distribution of $\epsilon$ in different TDEs.\\
    
     \item Stellar mass $\msr$.\\As introduced above, we adopt a Kroupa IMF, whose minimum and maximum stellar masses are 0.08 $\msun$ and 1 $\msun$, respectively.\\
     
     \item Scaled impact parameter $b$. It is a proxy for the impact parameter $\beta$.\\For $b$, we referred to the assumption of \citet{2020ApJ...890...73B}, that the probability for an encounter with a pericentre distance between $\rp$ and $\rp+\mrm{d}\rp$ is proportional to the area $2\pi\rp\,\mrm{d}\rp$, and the corresponding distribution function of $\beta$:
    \begin{equation}
    p(\beta)=\frac{1}{2\,\beta^3}\left(\frac{1}{\beta_{\mrm{min}}^2}-\frac{1}{\beta_{\mrm{max}}^2}\right)^{-1}.
    \end{equation}
    
    where 
    \begin{equation}
        \begin{aligned}
            &\beta_{\mrm{min}}=\beta(b=0),\ \beta_{\mrm{max}}=\mrm{min}[\beta(\rp=2\rs)],\\
            &\beta(\rp=2\rs)=11.8(\mbh/10^6\msun)^{-2/3}(\msr/\msun)^{7/15}.
        \end{aligned}
    \end{equation}
    
    The selection of $\beta(b=0)$ and $\beta$ will be explained below. Then we convert $\beta$ into $b$. According to the work of \citet{2013ApJ...767...25G} based on the polytropic models, for a star with $\gamma=4/3$, minimum disruptions correspond to $b=0$ and $\beta=0.6$, full disruptions correspond to $b=1$ and $\beta=1.85$, and disruptions with $b=2$ correspond to $\beta=4.0$. While for a star with $\gamma=5/3$, $b=0,1,2$ correspond to $\beta=0.5,0.9,2.5$, respectively. The {\tt MOSFiT} TDE model uses a hybrid polytropic model that blends between $\gamma=4/3$ and $\gamma=5/3$, in which a fraction of $\gamma=5/3$ is defined as 
    \begin{equation}
        f=\left\lbrace
        \begin{aligned}
            &1 ,& (0.08\leqslant\msr/\msun\leqslant0.3)\\
            &1-\frac{(\msr/\msun)-0.3}{1.0-0.3}  ,& (0.3<\msr/\msun\leqslant1.0)\\
        \end{aligned}\right.
    \end{equation}
    within the mass range from which we drew \citep{2019ApJ...872..151M}. Therefore, for a star with $0.08\leqslant\msr/\msun\leqslant1.0$, its
    \begin{equation}
    \begin{aligned}
    &\beta(b=0)=0.5f+0.6(1-f)=0.6-0.1f=\beta_{\mrm{min}},\\
    &\beta(b=1)=0.9f+1.8(1-f)=1.8-0.9f,\\
    &\beta(b=2)=2.5f+4.0(1-f)=2.5-1.5f.    
    \end{aligned}
    \end{equation}
    
    Using a piecewise linear function $b=b(\beta)$, and setting an upper limit of $b_{\mrm{max}}=2$, we finally finish the conversion from $\beta$ to $b$.
\end{enumerate}

\begin{figure}
    \centering
    \includegraphics[width=8.5cm]{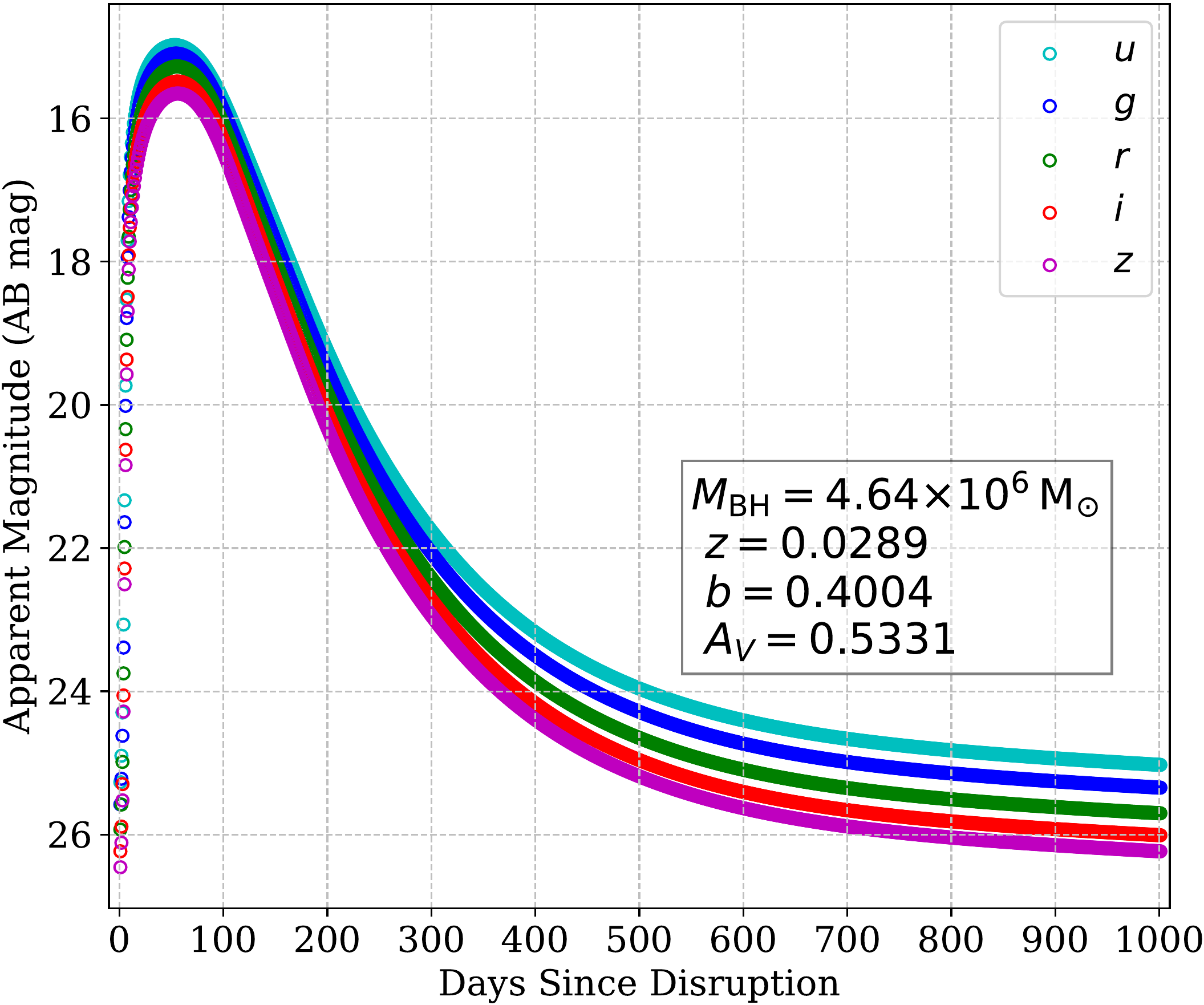}
    \caption{An example of TDF light curves at WFST $\ugriaz$ bands generated by {\tt MOSFiT} model. The light curves have been sampled from day 1 to 1001 since disruption with an interval of one day.}
    \label{fig:mosfitex}
\end{figure}

After setting these parameters, and convolving with the WFST filter throughput curves, the generated 
light curves of a TDE at $u,\ g,\ r,\ i,\ z$ bands are obtained (see an example in Figure~\ref{fig:mosfitex}). For each galaxy, we have drawn 10 pairs of $\msr$ and $b$, and then drawn 10 $A_V$ for each pair, and correspondingly, generated 100 light curves.

We have implemented a final selection before the mock observation. 
If the peak magnitudes of a TDE (TDF plus host) are below the WFST 30-second limiting
magnitudes, that are $m_{u}=22.31,\ m_{g}=22.93,\ m_{r}=22.77,\ m_{i}=22.05,\ m_{z}=21.02$ \citep{2018AcASn..59...22S}, we exclude the source out as it is beyond the detection threshold. This will get rid of $\sim90\%$ galaxies.




\section{Mock Observation and Detection} \label{sec:mocko}

After host galaxies and TDF light curves prepared, we introduce 
mock observations and detection criteria in this section. 
We assume a simple uniform survey strategy for the experimental
440 deg$^2$ CosmoDC2 field, which will be scanned with
30-second single exposures every 10 days at $u,g,r,i,z$ band, respectively.
In reality, a specific source can be only well observed during half-time in a year, thus we put "observation windows" spanning 180 days in the timeline. On the other hand, the clear night proportion, that is about 70\% defined as more than 4 hours of contiguous fully clear time \citep{2021Natur.596..353D}, has been also considered.

\subsection{Mock images}

We begin with generating mock images of host galaxies using the 
open-source simulation tool {\tt GalSim}~\citep{2015A&C....10..121R}. 
First, the pixel size is fixed to the one used by WFST CCD, that is 0.333\arcsec/pixel.
The galaxy surface brightness distribution at a given band, 
comprising an exponential disk and a Sérsic bulge, is derived from 
its morphological parameters and SED. 
TDF appears as a point source in the galaxy centre. 
However, the real images are much more complicated as they will be blurred by atmospheric seeing and contaminated by noises.



\subsubsection{Seeing}
The atmospheric seeing of Lenghu Site is being monitored and publicly available\footnote{http://lenghu.china-vo.org/sitecondition}. 
We blurred images at different epochs with Gaussian point spread functions (PSFs), whose full width at half maximum (FWHM) is equal to the seeing values at that epoch.
The assigned seeing at different epochs obeys the probability distribution from monitoring (median 0.78\arcsec).

\subsubsection{Noise}
We consider two origins of noises. One is the Poisson noise and the other is the readout noise.  First, the Poisson noise obeys the Poisson distribution,
\begin{equation}
P(k,N)=\frac{N^k}{k!}e^{-N},\ k\in\mathbb{N}_0,
\end{equation}
where $N$ is the total count from astronomical sources and sky background in each pixel. The readout noise is subject to a standard normal distribution: $X\sim \mathcal{N}(\mu,\sigma^2)$, where $\mu=0$, and $\sigma$ is set to $10\,e^{-}/$pixel. 

\subsection{Reference Images}

In addition to mock images for a single visit, we further need to construct 
reference images  since almost all modern time-domain surveys 
have chosen to detect transients in the reference-subtracted difference images directly or by comparing with the photometry on reference images.
The reference images are normally created by stacking high quality images to
a depth deeper than individual images. 
Here we simply define the images observed under conditions of 
seeing $\leqslant$ 0.7" as being good enough to be stacked. 
If the first-year survey has uniformly scanned the whole northern sky at all five bands, we can obtain reference images identical to a seeing of
0.7" and an exposure time of 100 seconds in each band.

\begin{figure*}
    \centering
    \begin{subfigure}{\textwidth}
        \includegraphics[width=\textwidth]{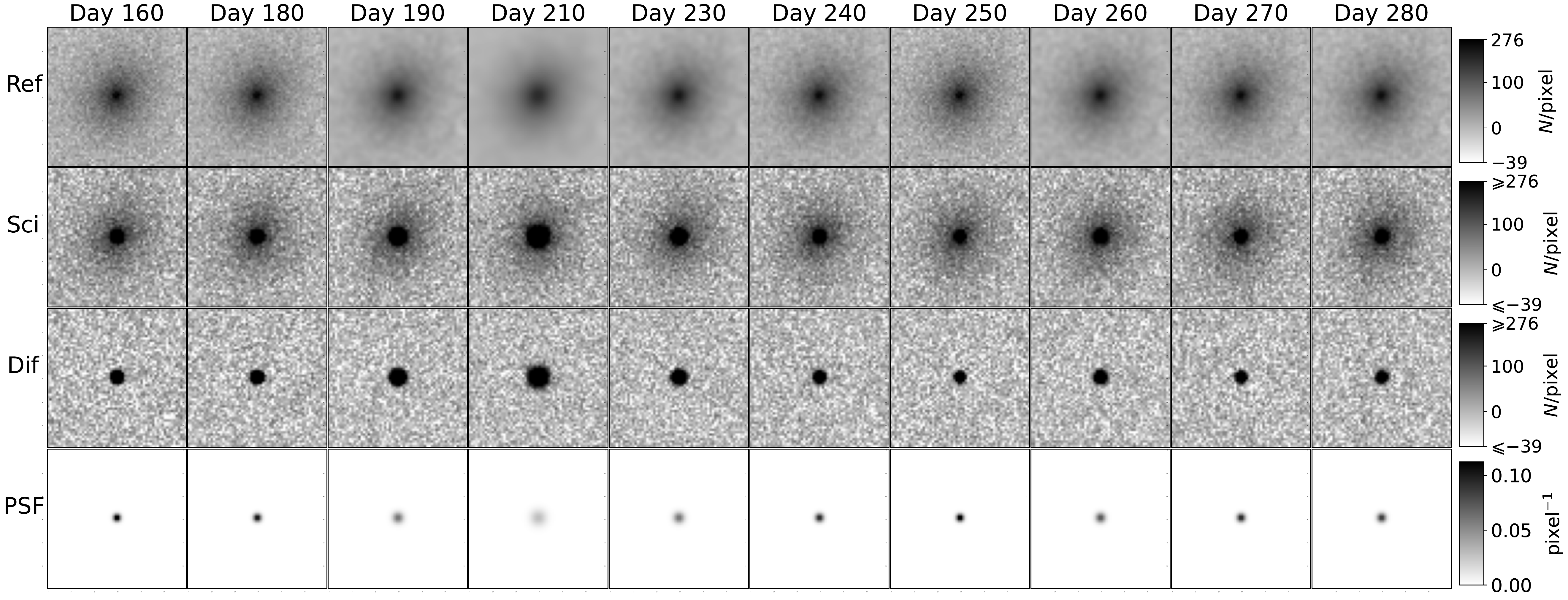}        
    \end{subfigure}
    \begin{subfigure}{\textwidth}
        \includegraphics[width=\textwidth]{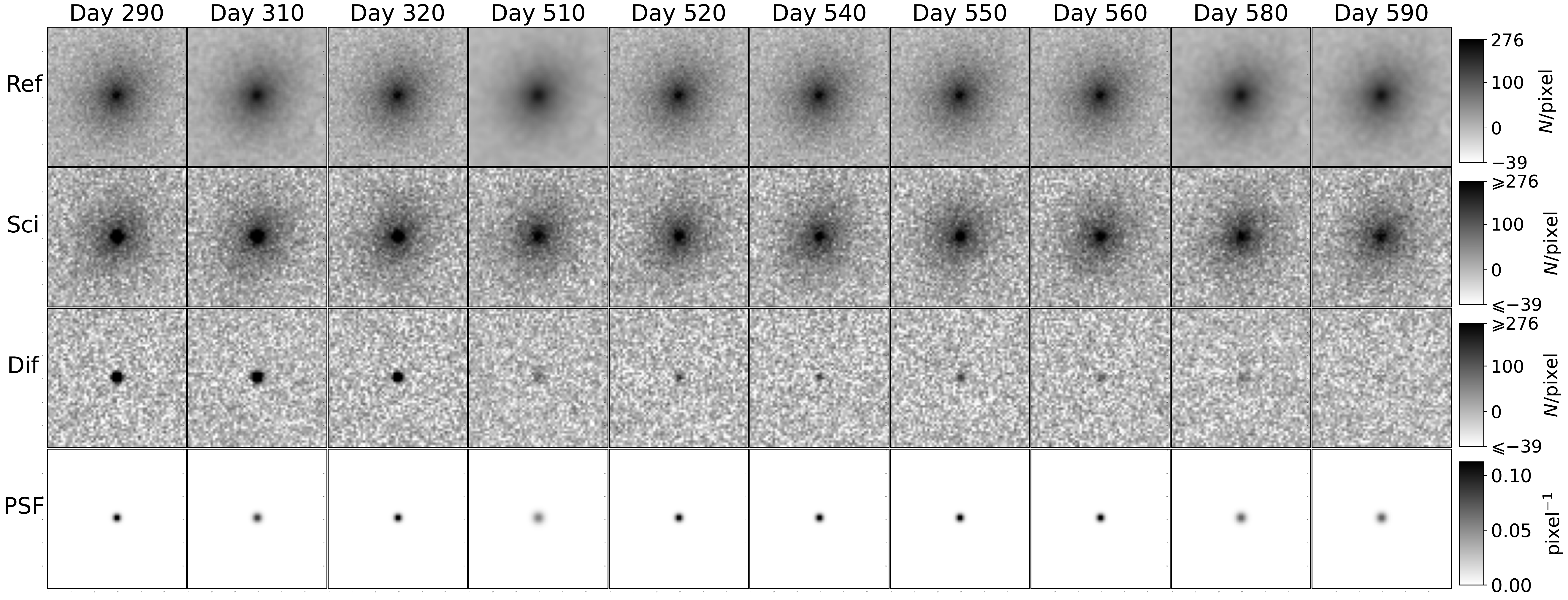}        
    \end{subfigure}
    \caption{The reference, science, difference and PSF images (from top to bottom) at different stages for a typical detectable TDE at $z=0.0452$. "Observation windows" have been put onto the timeline. Each "window" lasts for 180 days, and the intervals are also set as 180 days. Besides, a clear night proportion for WFST, about 70\%, is also considered, making the final timeline. The seeing of original reference images is 0.7\arcsec. The reference images are convoluted to match the atmospheric seeing of the science images if the seeing is worse than 0.7\arcsec (e.g., Day 210), and vice versa (e.g., Day 250). We note that the difference images are still far from reality, therefore we performed photometry only on reference and science images.}
    \label{fig:timeline}
\end{figure*}

\subsection{Light curves from PSF photometry} \label{sec:psf}

The PSF photometry has been widely applied for transient detection by
time-domain surveys nowadays, either in difference images or original images, 
because it can effectively reduce the impact from starlight for detecting 
transients embedded in galaxies, such as TDEs in galaxy centres. 
The specific PSF photometry code we used is 
{\tt PythonPhot} \citep{2015ascl.soft01010J} which has also
considered the noise images well.
For each observation (in a given filter on a certain day), the PSF image
is produced by {\tt getpsf} function assuming its FWHM 
is approximately identical with seeing.

\begin{figure*}
    \centering
    \begin{subfigure}{0.48\textwidth}
        \centering
        \includegraphics[width=\textwidth]{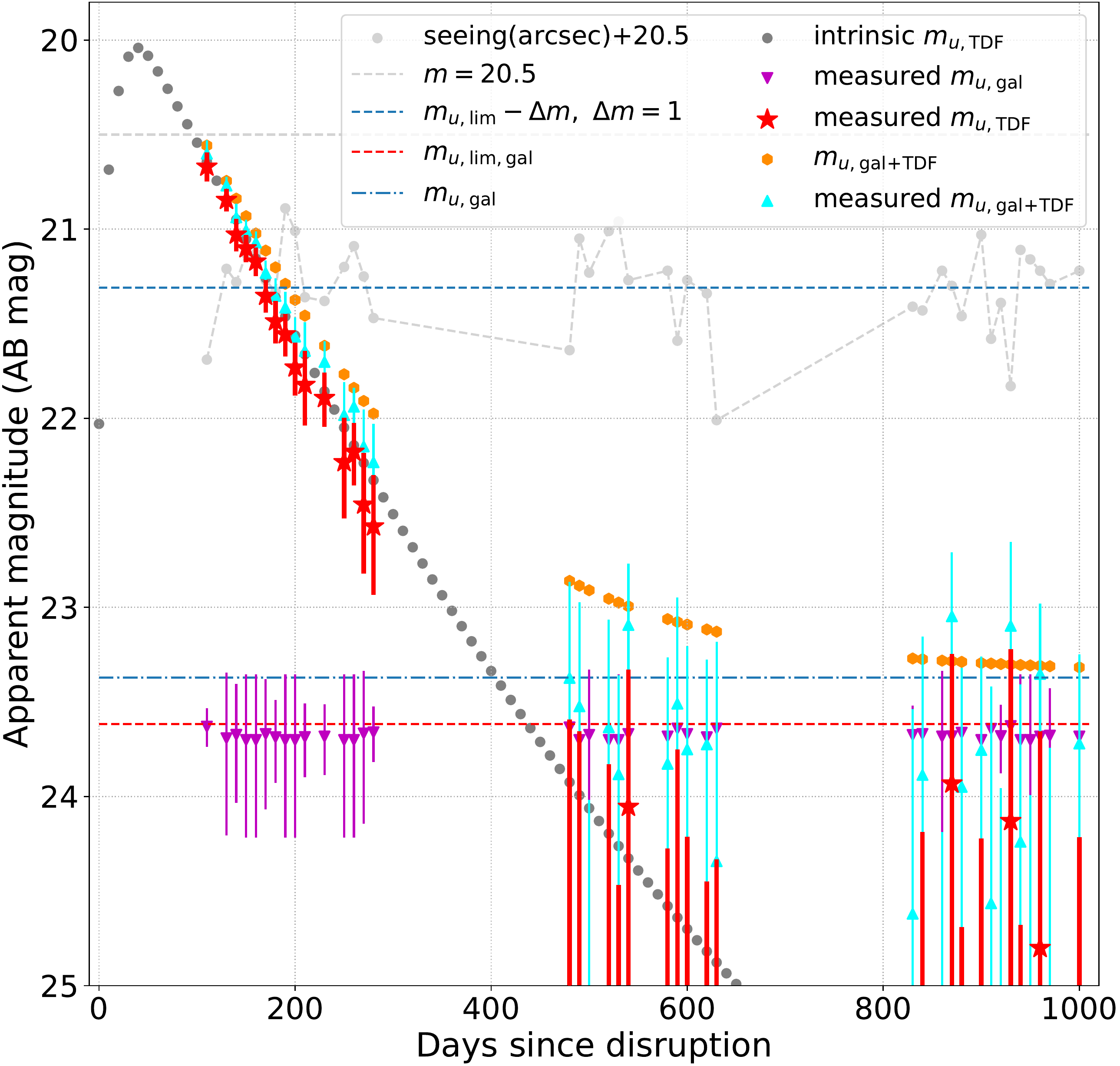}
        \caption{$u$ band}
    \end{subfigure}
    \begin{subfigure}{0.48\textwidth}
        \centering
        \includegraphics[width=\textwidth]{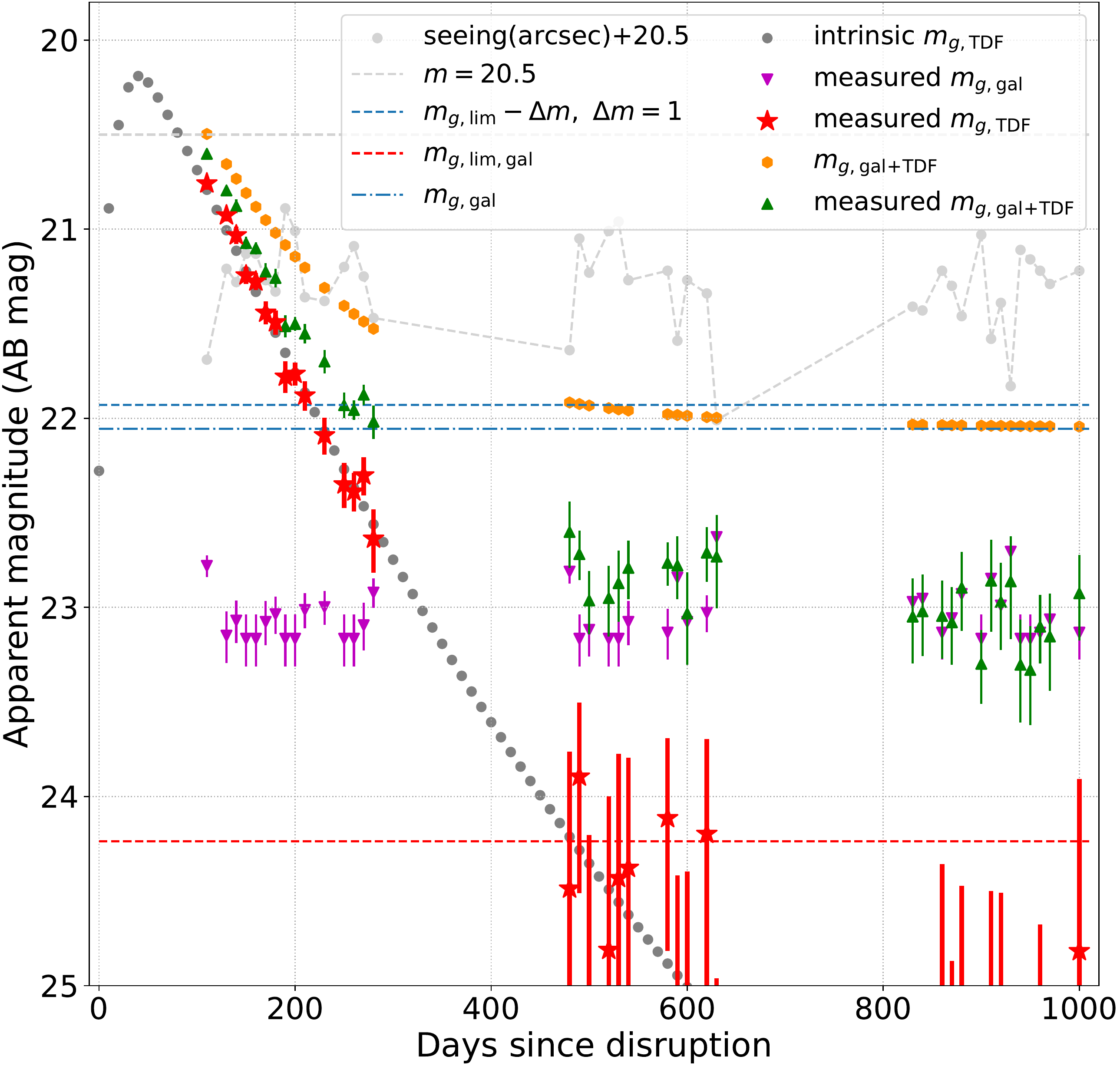}  
        \caption{$g$ band}
    \end{subfigure}
    \caption{The light curves of a TDE at $z=0.253$ as an example 
    of our mock observations. The left and right panel show $u$ and $g$ band results, respectively. The measured TDF light curves with 1$\sigma$ errors are shown in red, 
    which is calculated as the difference between the PSF photometry on the reference image and single-epoch observed image. 
    For the ease of comparison, we have also plotted the intrinsic galaxy magnitude 
    (blue dashdotted lines), intrinsic TDF magnitude given by MOSFiT (dark grey dots), 30-second (for science images, blue dashed) and 100-second (for reference images, red dashed) limiting magnitude in the corresponding band. 
    The seeing value at each epoch is also shown in light grey.
   \label{fig:psfph}}
\end{figure*}

We take the difference of PSF photometry between a single-visit and reference image as the contribution from TDF. Before this step, we have convolved the images with better seeing with Gaussian function to ensure they share the same imaging quality.
The PSF photometry is principally self-consistent and precise only for point sources.
Nevertheless, the TDF emission overlapped with its extended host galaxy is 
certainly not a strict point source. To check the systematic errors, we have
compared the derived TDF emission from our strategy with the one predicted 
from {\tt MOSFiT} model. It suggests that they agree with each other at most epochs, albeit
with large errors at late stages when TDF emission becomes very weak and thus being overwhelmed by starlight (see Figure~\ref{fig:psfph}). 
Generally, the measurement from the difference of PSF photometry is 
acceptable for our simulation.

\subsection{Criteria for TDE detection}\label{sec:crit}

After acquiring the TDE light curves by our measurements (as shown in Figure~\ref{fig:psfph}), we can then check and define the detection criteria for a TDE.
We use two key parameters, that is the difference as measured from PSF photometry ($\Delta N=N_{\mrm{gal+TDF,30s}}-N_{\mrm{gal,30s}}$) and their errors ($\sigma_{N}=\sqrt{(\Delta N_{\mrm{gal+TDF,30s}})^2+(\Delta N_{\mrm{gal,30s}})^2}$). 

A host galaxy detected in a band of WFST should satisfy this condition
\begin{equation}\label{eqn:con1}
\frac{N_{\mrm{gal,100s}}-N_{\mrm{lim,100s}}}{\Delta N_{\mrm{gal,100s}}}>s_1,\\
\end{equation}
where the limiting count $N_{\mrm{lim,100s}}$ stands for the counts of a 100-second exposure to a source having an apparent magnitude that equals to the 5-$\sigma$ limiting magnitude, $m_{\mrm{lim,100s}}\approx m_{\mrm{lim,30s}}+1.25\,\mrm{log}_{10}(10/3)$, and a TDE detected in a band of WFST should satisfy these two conditions
\begin{gather}
\mgaltdf<m_{\mrm{lim,30s}}-\Delta m;\label{eqn:con2}
\\
\frac{\Delta N}{\sigma_N}>s_2\label{eqn:con3}
\end{gather}
for $k$ times.

For $\ugriaz$ band, a detectable TDE should be detected in at least $l_1$ band(s), with its host galaxy detected in at least $l_2$ band(s). We set $s_1=s_2=3$, $\Delta m=1$, $k=10$, $l_1=2$, $l_2=1$. As mentioned above, $m_{\mrm{lim,30s}}$ of five WFST bands are $m_{u}=22.31,\ m_{g}=22.93,\ m_{r}=22.77,\ m_{i}=22.05,\ m_{z}=21.02$.

\section{Results} \label{sec:results}

In this section, we will present the results from our mock observations
and try to analyse the impact by different factors, e.g., band selection, noise and sampling cadence.

\subsection{Noise \& Band combination}

\begin{figure*}
    \centering
    \begin{subfigure}{0.33\textwidth}
        \centering
        \includegraphics[width=\textwidth]{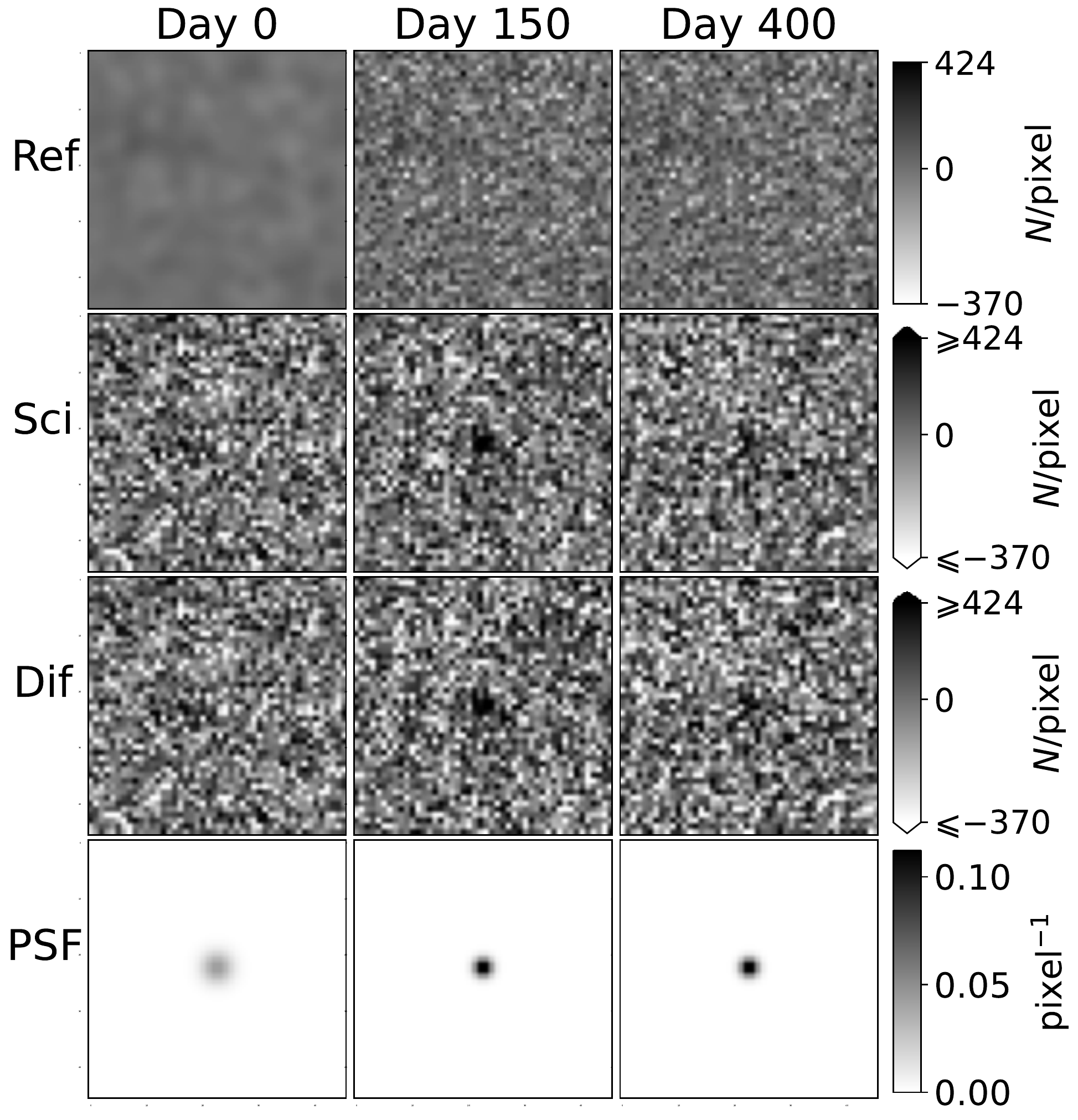}
        \caption{$I_{\mrm{sky}}=17$ mag/arcsec$^2$, $\sigma=0\,e^{-}/$pixel}
    \end{subfigure}
    \begin{subfigure}{0.33\textwidth}
        \centering
        \includegraphics[width=\textwidth]{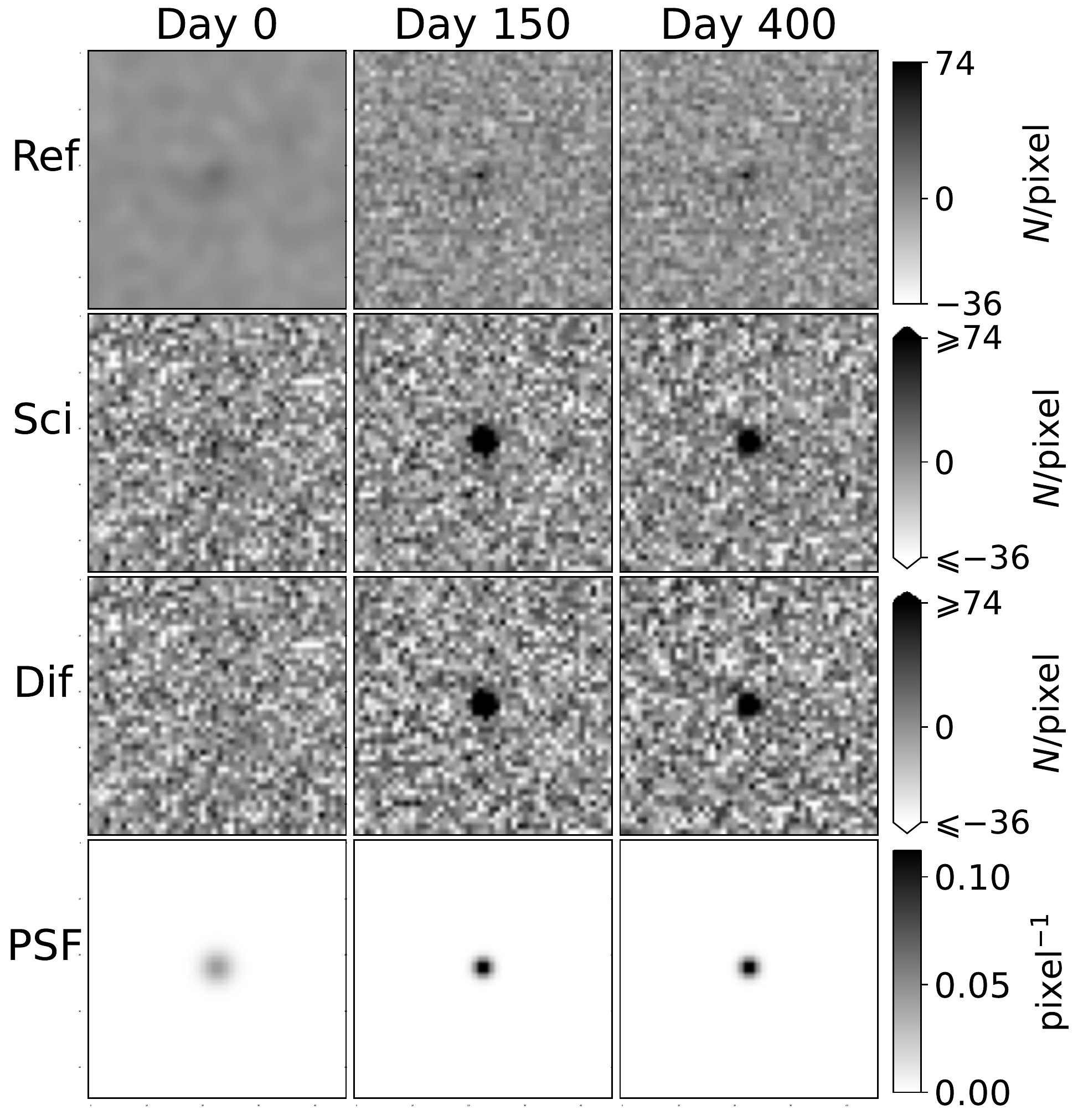}
        \caption{$I_{\mrm{sky}}=22$ mag/arcsec$^2$, $\sigma=0\,e^{-}/$pixel}
    \end{subfigure}
    \begin{subfigure}{0.33\textwidth}
        \centering
        \includegraphics[width=\textwidth]{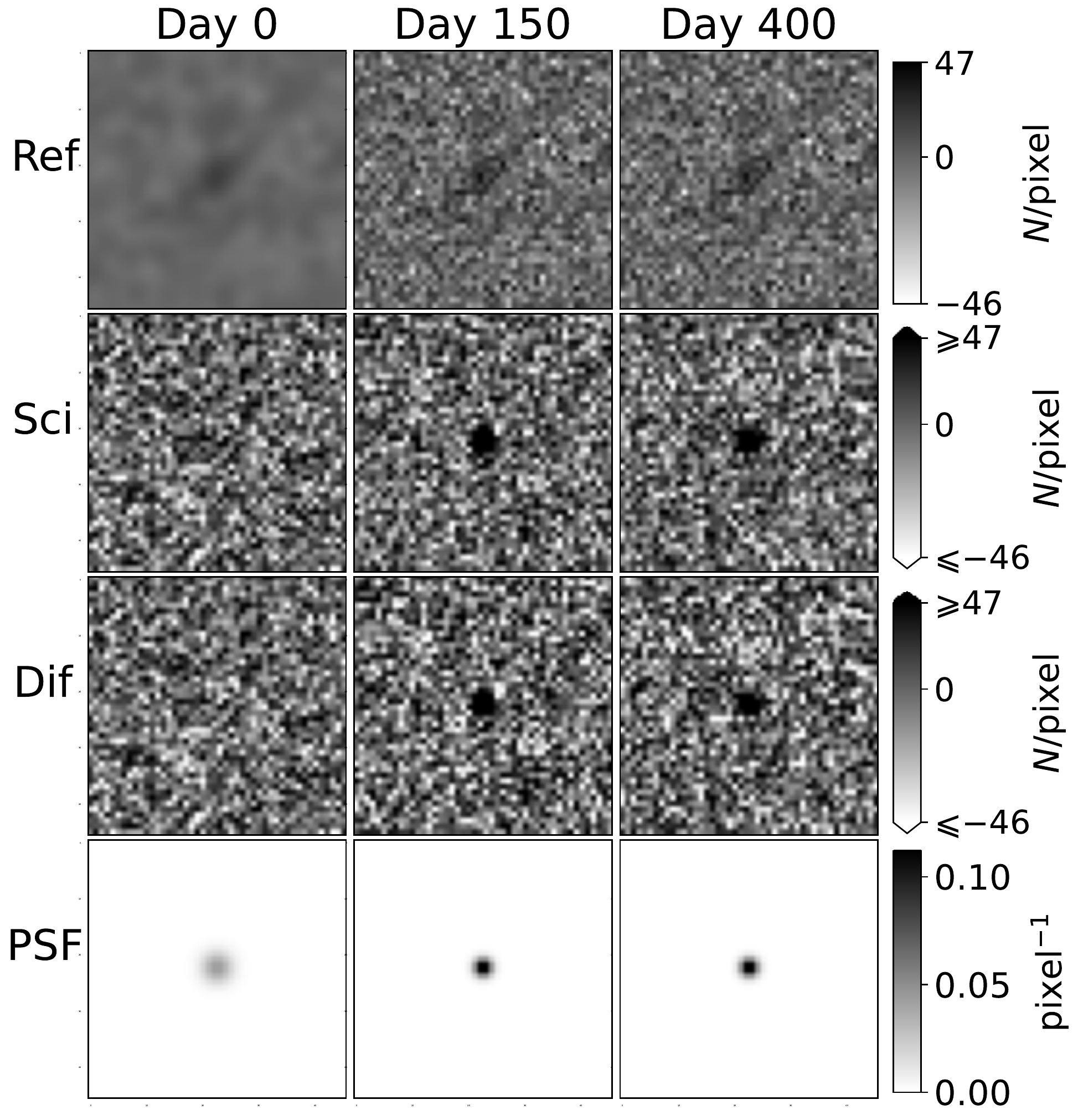}
        \caption{$I_{\mrm{sky}}=22$ mag/arcsec$^2$, $\sigma=10\,e^{-}/$pixel}
    \end{subfigure}
    \caption{Illustrations for the impact of sky background and readout noise. 
    Panel (a) shows the reference, science, difference and PSF images (from top to bottom) for a TDE observed with $I_{\mrm{sky}}=17$ mag/arcsec$^2$, $\sigma=0\,e^{-}/$pixel, i.e., no readout noise. (b): $I_{\mrm{sky}}=22$ mag/arcsec$^2$, $\sigma=0\,e^{-}/$pixel. (c): $I_{\mrm{sky}}=22$ mag/arcsec$^2$, $\sigma=10\,e^{-}/$pixel.
    \label{fig:noise}}
\end{figure*}

Lenghu Site has sky background monitoring data in the Sky Quality Metre (SQM) photometric system \citep{2005ISTIL...9...1}. However, the conversion factors from SQM readings to WFST bands are quite uncertain, so we just simply estimate the influence by setting several typical sky background values. We have tested six levels of sky backgrounds, that are $I_{\mrm{sky}}=17,18,19,20,21,22$ mag/arcsec$^2$. Likewise, the readout noise has been also taken into account.
The impact of noise to the signal detection can be easily seen in Figure~\ref{fig:noise}.

We have also checked the affection of noise to observations with
different band selections, which is presented in Table~\ref{tab:noise}.
We may immediately conclude from them as below.

\begin{table*}
\begin{threeparttable}
\centering
\caption{Results of 10$\times$10$\times$10 mock observations: The impact of sky background and readout noise\label{tab:noise}}
\begin{tabular}{c|c|ccccccc}
\hline
\multicolumn{2}{c|}{$I_{\mrm{sky}}$(mag/arcsec$^2$)} & 17      & 18      & 19      & 20      & 21      & 22      & 22      \\ \hline
\multicolumn{2}{c|}{Readout noise($e^-$/pixel)}                   & 10     & 10     & 10     & 10     & 10     & 10     & 0      \\ \hline
                         & $ug$                      & 3(74)  & 5(128)  & 7(187)  & 9(237)  & 10(264) & 11(290) & 11(299) \\
                         & $ur$                      & 5(124) & 8(226)  & 12(327) & 15(408) & 17(449) & 17(460) & 18(481) \\
                         & $ui$                      & 3(92)  & 6(172)  & 10(262) & 12(330) & 13(359) & 14(373) & 15(389) \\
                         & $gr$                      & 5(146) & 10(275) & 16(437) & 23(610) & 27(712) & 28(757) & 30(804) \\
                         & $gi$                      & 4(99)  & 7(187)  & 11(292) & 14(378) & 16(419) & 17(445) & 17(467) \\
Used Bands               & $ri$                      & 4(107) & 8(207)  & 12(317) & 15(408) & 17(446) & 17(466) & 18(487) \\
                         & $ugr$                     & 6(106) & 11(193) & 17(301) & 23(415) & 27(482) & 29(512) & 30(543) \\
                         & $ugi$                     & 5(98)  & 9(166)  & 13(239) & 16(290) & 18(322) & 19(339) & 20(359) \\
                         & $uri$                     & 5(97)  & 10(173) & 14(252) & 17(312) & 19(344) & 20(358) & 21(378) \\
                         & $gri$                     & 6(113) & 12(208) & 18(315) & 24(428) & 27(488) & 29(515) & 31(548) \\
                         & $ugri$                    & 7(91)  & 12(163) & 18(244) & 24(327) & 28(372) & 29(392) & 31(417) \\
                         & $ugriz$                   & 7(74)  & 12(130) & 18(195) & 24(261) & 28(297) & 29(314) & 31(333)  \\\hline
\end{tabular}
\begin{tablenotes}
\item \textbf{Notes:}
\item 1. Out of parentheses: Detectable TDEs in 440 deg$^{2}$ per year, with sky background level, readout noise and band combination altered. 
\item 2. In parentheses: Detectable TDEs in maximum survey area per year, with sky background level, readout noise and band combination altered. The maximum survey area are calculated based on the WFST scanning speed (assuming 4-hour useful observing time per night, 10-second readout time), with an upper limit of 2$\pi$ steradian ($\sim20,600$ deg$^2$) if the band combination can fully cover the full northern celestial sphere.
\end{tablenotes}
\end{threeparttable}
\end{table*}

\begin{figure*}
    \centering
    \begin{subfigure}{0.45\textwidth}
        \centering
        \includegraphics[width=\textwidth]{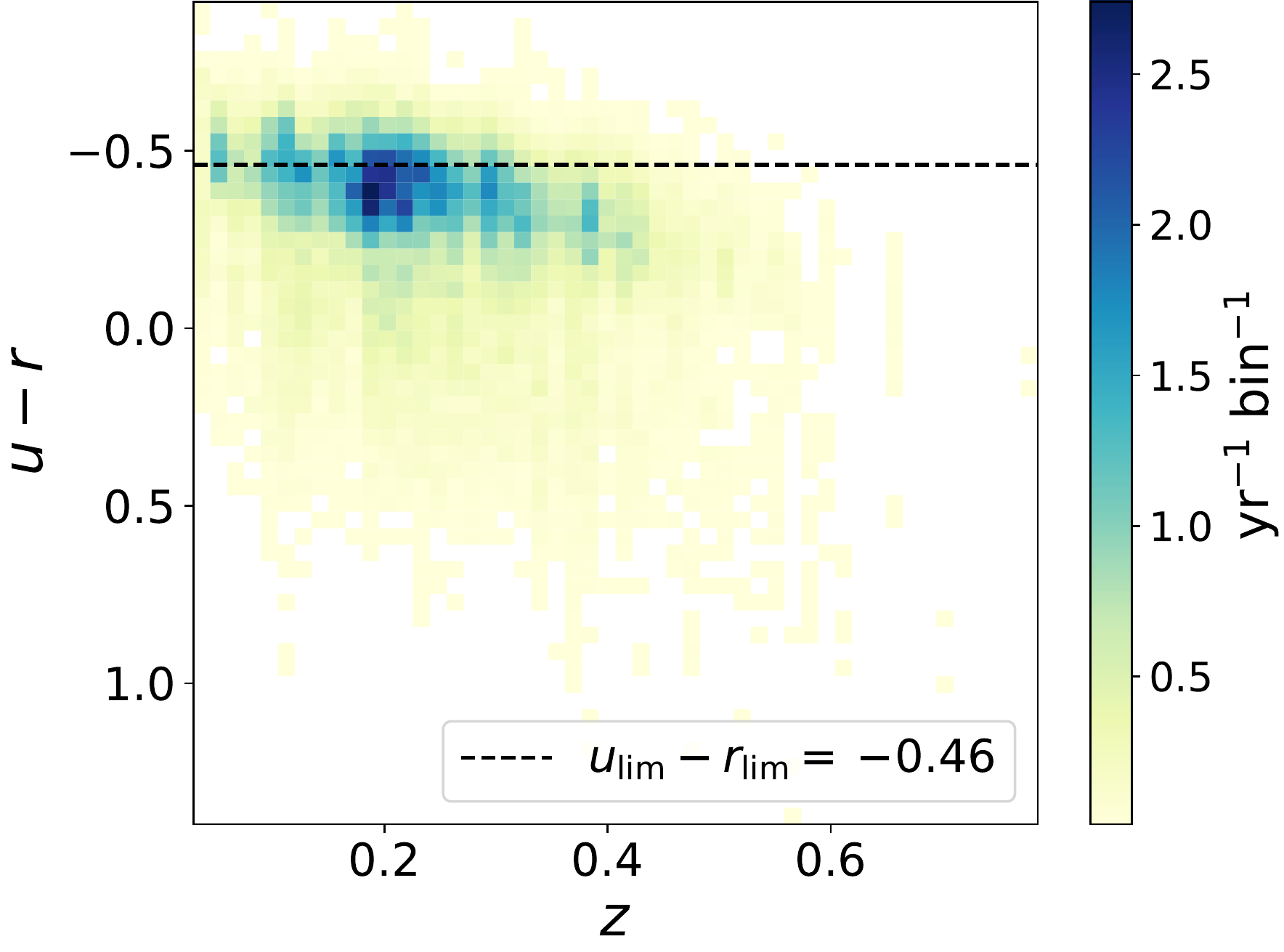}
        \caption{TDF}
    \end{subfigure}
    \begin{subfigure}{0.45\textwidth}
        \centering
        \includegraphics[width=\textwidth]{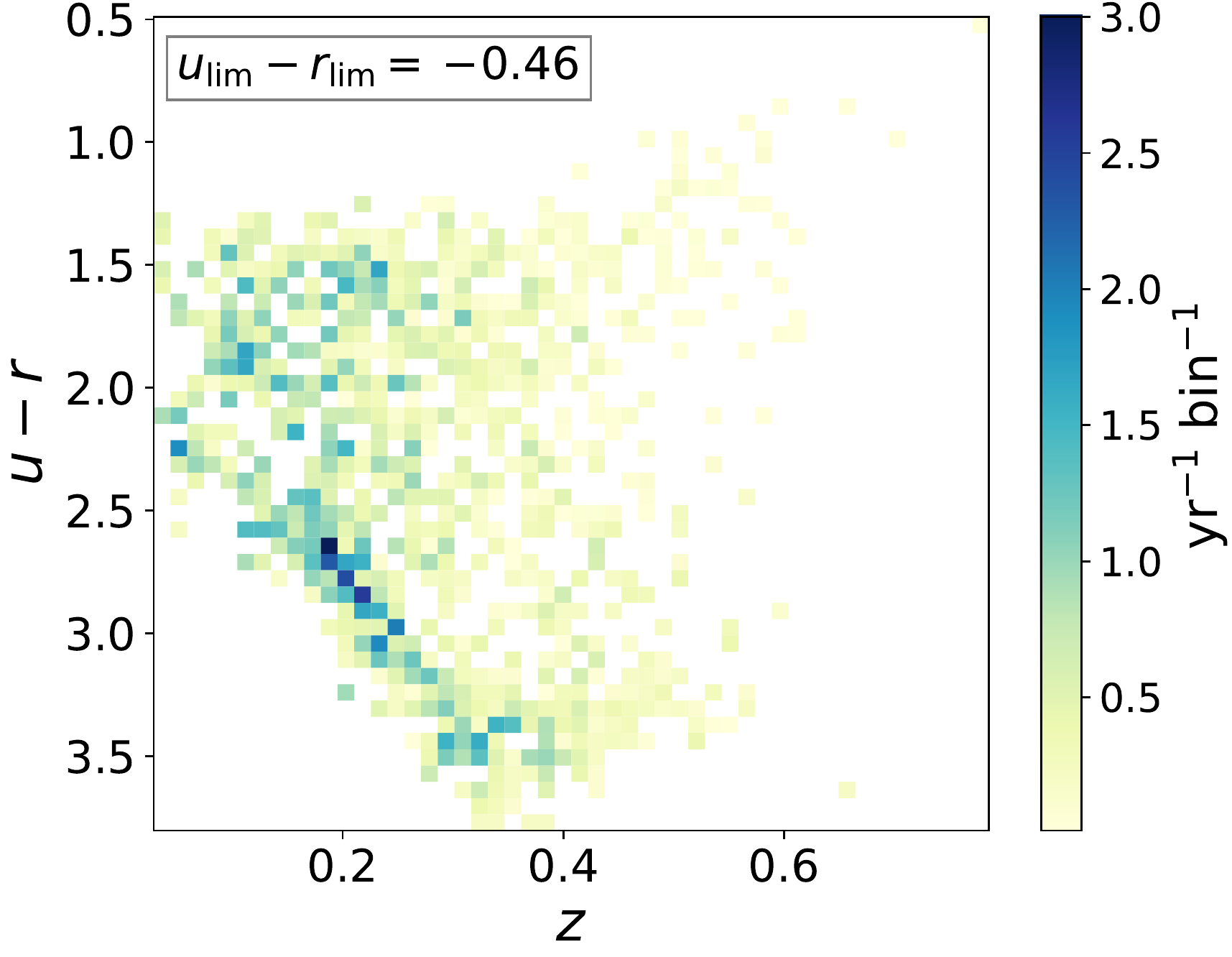}
        \caption{Host galaxy}
    \end{subfigure}\\
    \begin{subfigure}{0.45\textwidth}
        \centering
        \includegraphics[width=\textwidth]{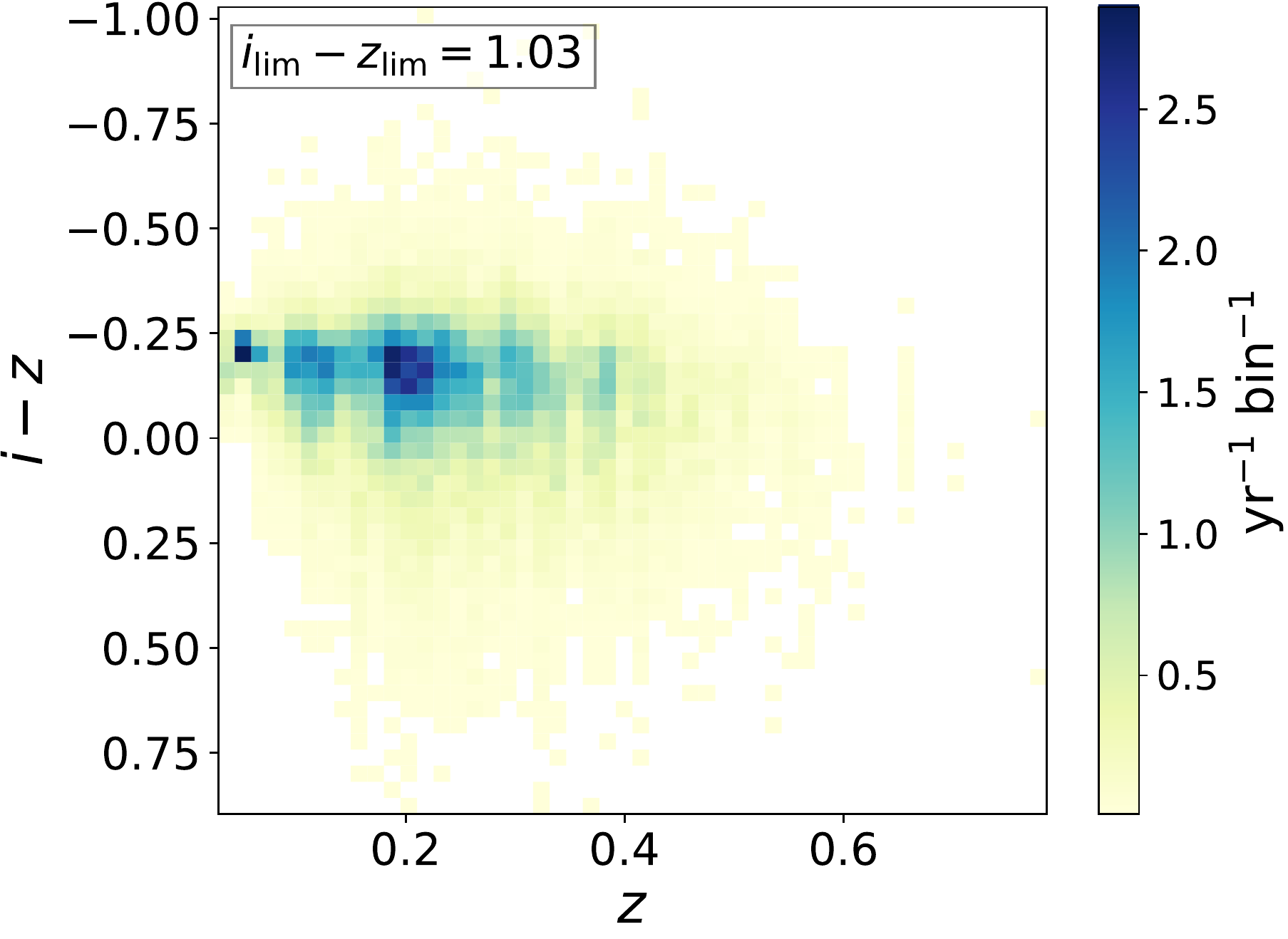}
        \caption{TDF}
    \end{subfigure}
    \begin{subfigure}{0.45\textwidth}
        \centering
        \includegraphics[width=\textwidth]{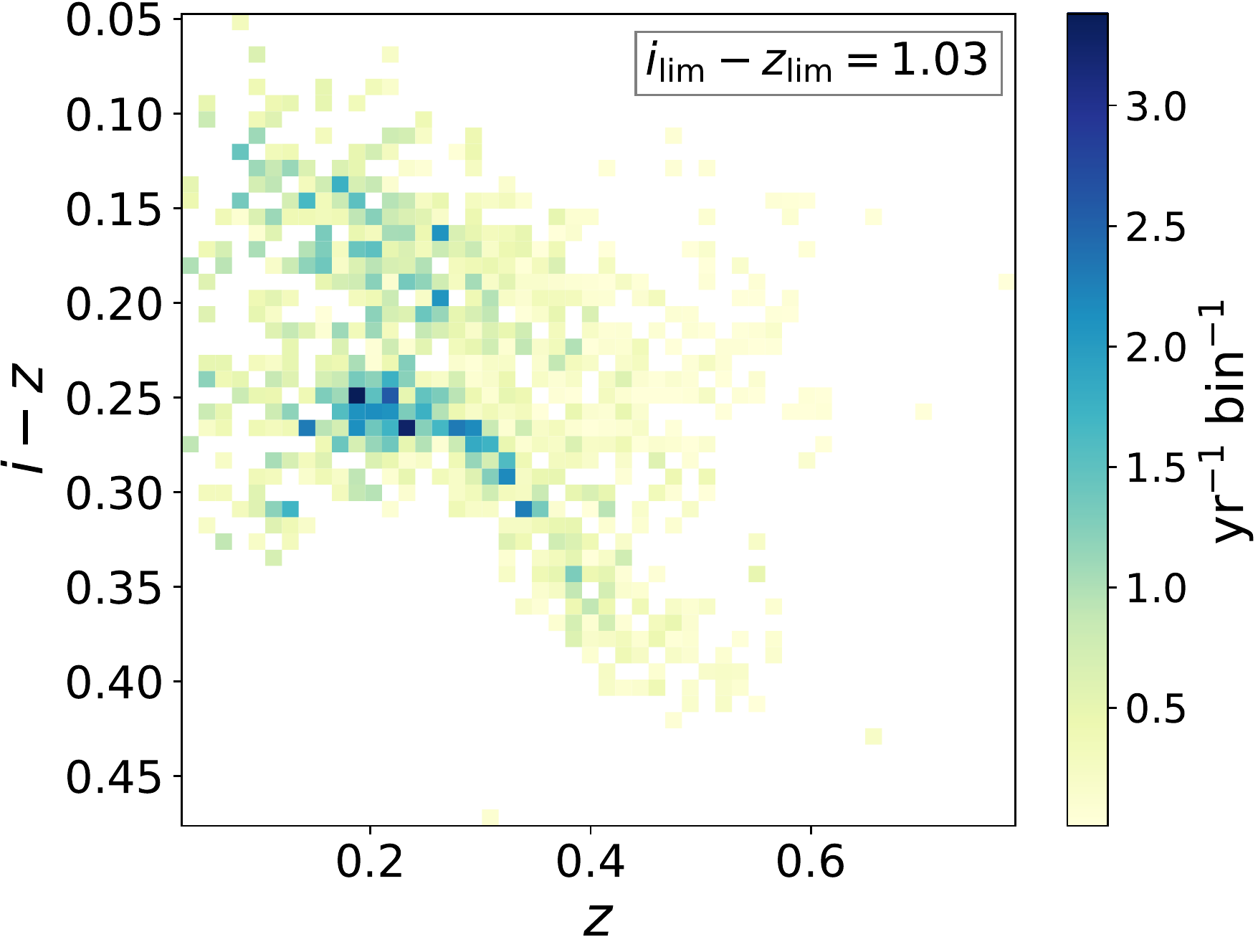}
        \caption{Host galaxy}
    \end{subfigure}
    \caption{Top panels: the $u-r$ colour of detectable TDFs (left) and host galaxies (right). For most TDFs and almost all host galaxies, the colour is redder than the difference between the limiting magnitude of $u$ and $r$ band, which is $u_{\mrm{lim}}-r_{\mrm{lim}}=-0.46$ (dashed horizontal line). As a result, $u$ band cannot significantly increase the detection rate of TDE when both $g$ and $r$ band have been included in the observations.
    Bottom panels: similar to top panels yet $i-z$ colour has been shown. The difference between the limiting magnitude of $i$ and $z$ band is $i_{\mrm{lim}}-z_{\mrm{lim}}=1.03$, that is redder than the $i-z$ colour of TDFs and host galaxies. Therefore, $z$ band contributes little to the TDE detection.
    \label{fig:uriz}}
\end{figure*}

\begin{enumerate}
    \item The sky background affects the results significantly when $I_{\mrm{sky}}<20$ mag/arcsec$^2$.
    \item The readout noise slightly influences the whole detection rate with an exposure time of 30 seconds for single visit. It not only contaminates the TDF emission, but also impedes the detection of host galaxies, especially at high redshifts.
    \item $g$ and $r$ are the most crucial bands for TDE detection.
    \item $u$ band cannot significantly increase the detection rate directly once both $g$ and $r$ bands have been adopted. The uselessness for detection is mainly due to the shallow magnitude limit in $u$ band. The difference between the limiting magnitude of $u$ and $r$ band is $u_{\mrm{lim}}-r_{\mrm{lim}}=-0.46$,
    which is smaller than that of galaxies and most TDFs, as shown in top panels of Figure \ref{fig:uriz}. However, $u$ band should play a key role in the SED fitting of TDEs as most optical-UV TDEs have the blackbody temperature $T_{\mrm{bb}}$ in the range of 10,000 to 50,000 Kelvin \citep{2020SSRv..216..124V}, corresponding to the rest-frame peak wavelength range of $580$ to $2900$ \AA. Therefore, the $u$-band, which is closest to the SED peak among all filters, 
    would be very useful to constrain the SED in the practical transient classification. For this reason, we will still recommend a selection of $u$
    band in the TDE search.
    \item $z$ band cannot significantly improve the detection rate of TDEs, since the difference between the limiting magnitude of $i$ and $z$ band is $i_{\mrm{lim}}-z_{\mrm{lim}}=1.03$, which is larger than the $i-z$ of TDFs and host galaxies (see bottom panels of Figure~\ref{fig:uriz}). Moreover, it can hardly put effective constraints on the SED fitting, that is distinct from $u$ band. Therefore, we will abandon $z$ band in the considerations of filters.
\end{enumerate}

\subsection{Threshold \& Cadence}

Mock observations were then performed on the first 10 groups of galaxies, with 100 observations per group. In default, we set threshold $k=10$, cadence $C=10$ d, as introduced in Section \ref{sec:crit}. Changing either of them might significantly affect the results, hence we set $k=5$ and $C=20$ d. We have kept $s_1=s_2=3$, $\Delta m=1$, $l_1=2$, and $I_{\mrm{sky}}=22$ mag/arcsec$^2$. The results are shown in Table \ref{TAB:THRCAD}. WFST can cover the full northern celestial sphere within a $\sim$35-hour observing period in a single filter. Consider an average exposure of 4 hours per night, we note that even $C=20$ d cannot cover the full northern celestial sphere in more than two filters.

Table~$\ref{TAB:THRCAD}$ indicates that higher cadence and lower threshold will
result in higher detection rates, in agreement with our expectation.

\subsection{Host galaxy detection}\label{sec:host}

The host galaxy detection is crucial for the identification of TDE candidates
from numerous of unclassified transients because the isolated or off-nuclear 
transients, which are vast majority, can be cast away with the aid of the association 
with host galaxy. Deeper reference images or those taken from other
value-added catalogues could greatly benefit the host detection. For the former one, the results based on 100-second (i.e., one-year survey on the northern sky) reference images have been shown above, while the latter is beyond the scope of our work.

Additionally, the characteristics of host galaxies (e.g., colour) might also provide useful clues to transient 
classification before any follow-up observations. For this reason, the host galaxy
is better to be detected in more than two bands, i.e., $l_2=2$. We have checked its
affection and displayed main results in Table~\ref{TAB:THRCAD}, with $s_1=s_2=3$, $\Delta m=1$, $l_1=2$ retained. More details are shown in appendix (see Table~\ref{tab:full}).

\begin{table*}
\begin{threeparttable}
\centering
\caption{Results of 10$\times$10$\times$10 mock observations: The impact of threshold, cadence and $l_2$\label{TAB:THRCAD}}
\begin{tabular}{c|c|cccccccc}
\hline
\multicolumn{2}{c|}{Threshold $k$}  & 10  & 10  & 5   & 5   & 10  & 10  & 5   & 5  \\ \hline
\multicolumn{2}{c|}{Cadence $C$(d)} & 10  & 20  & 10  & 20  & 10  & 20  & 10  & 20 \\ \hline
\multicolumn{2}{c|}{$l_2$}          & 1   & 1   & 1   & 1   & 2   & 2   & 2   & 2  \\ \hline
& $ug$ & 11(290) & 4(183) & 16(428) & 11(517) & 1(32) & 0(23) & 2(46) & 1(58)\\
& $ur$ & 17(460) & 5(233) & 27(732) & 18(833) & 1(26) & 0(15) & 1(39) & 1(45)\\
& $ui$ & 14(373) & 4(172) & 23(612) & 14(674) & 1(21) & 0(10) & 1(33) & 1(38)\\
& $gr$ & 28(757) & 10(479) & 41(1108) & 29(1362) & 15(400) & 6(292) & 21(558) & 15(715)\\
& $gi$ & 17(445) & 5(219) & 27(713) & 17(807) & 11(294) & 4(168) & 17(446) & 11(530)\\
& $ri$ & 17(466) & 5(224) & 28(754) & 18(846) & 17(443) & 5(218) & 26(705) & 17(801)\\
& $ugr$ & 29(512) & 11(378) & 42(747) & 30(1054) & 15(274) & 7(236) & 21(381) & 16(561)\\
& $ugi$ & 19(340) & 6(213) & 30(528) & 20(702) & 12(221) & 5(162) & 18(323) & 13(453)\\
& $uri$ & 20(358) & 6(213) & 32(563) & 21(743) & 18(328) & 6(199) & 29(511) & 19(677)\\
& $gri$ & 29(515) & 10(369) & 42(756) & 30(1060) & 25(455) & 9(334) & 37(661) & 26(936)\\
& $ugri$ & 29(392) & 11(286) & 43(573) & 30(807) & 26(346) & 10(260) & 38(503) & 27(714)\\
& $ugriz$ & 29(314) & 11(229) & 43(459) & 30(646) & 26(277) & 10(208) & 38(402) & 27(571)\\
\hline
\end{tabular}
\begin{tablenotes}
\item \textbf{Notes:}
\item 1. Out of parentheses: \\Detectable TDEs in 440 deg$^{2}$ per year, with threshold, cadence, the minimum number of WFST bands that can detect the host galaxy and band combination altered.
\item 2. In parentheses: \\Detectable TDEs in maximum survey area per year, with threshold, cadence, the minimum number of WFST bands that can detect the host galaxy and band combination altered. The maximum survey area are calculated based on the WFST scanning speed (assuming 4-hour useful observing time per night, 10-second readout time), with an upper limit of 2$\pi$ steradian ($\sim$20,600 deg$^2$) if the band combination can fully cover the full northern celestial sphere. A detailed version of this table is displayed in appendix (see Table~\ref{tab:full}).
\item 3. Explanations on key parameters:\\
Threshold $k$: The minimum times that the TDF should be detected in a band of WFST. \\
Cadence $C$: The interval between two closest observations in any band of WFST.\\
$l_2$: The minimum number of WFST bands that can detect the host galaxy.
\end{tablenotes}
\end{threeparttable}
\end{table*}

\subsection{Comparison with real optical TDE sample}\label{sec:comp}

In this section, we will show the distribution of TDEs detected in our mock observations, and compare them with the 33 optical TDEs summarized in \citet{2020SSRv..216..124V}. Parameters set in these mock observations are: $s_1=s_2=3$, $\Delta m=1$, $k=10$, $l_1=2$, $l_2=1$, $C=10$ d, $ugri$ bands.

\subsubsection{Redshift and $\mbh$ distribution}

\begin{figure}
    \centering
    \includegraphics[width=8cm]{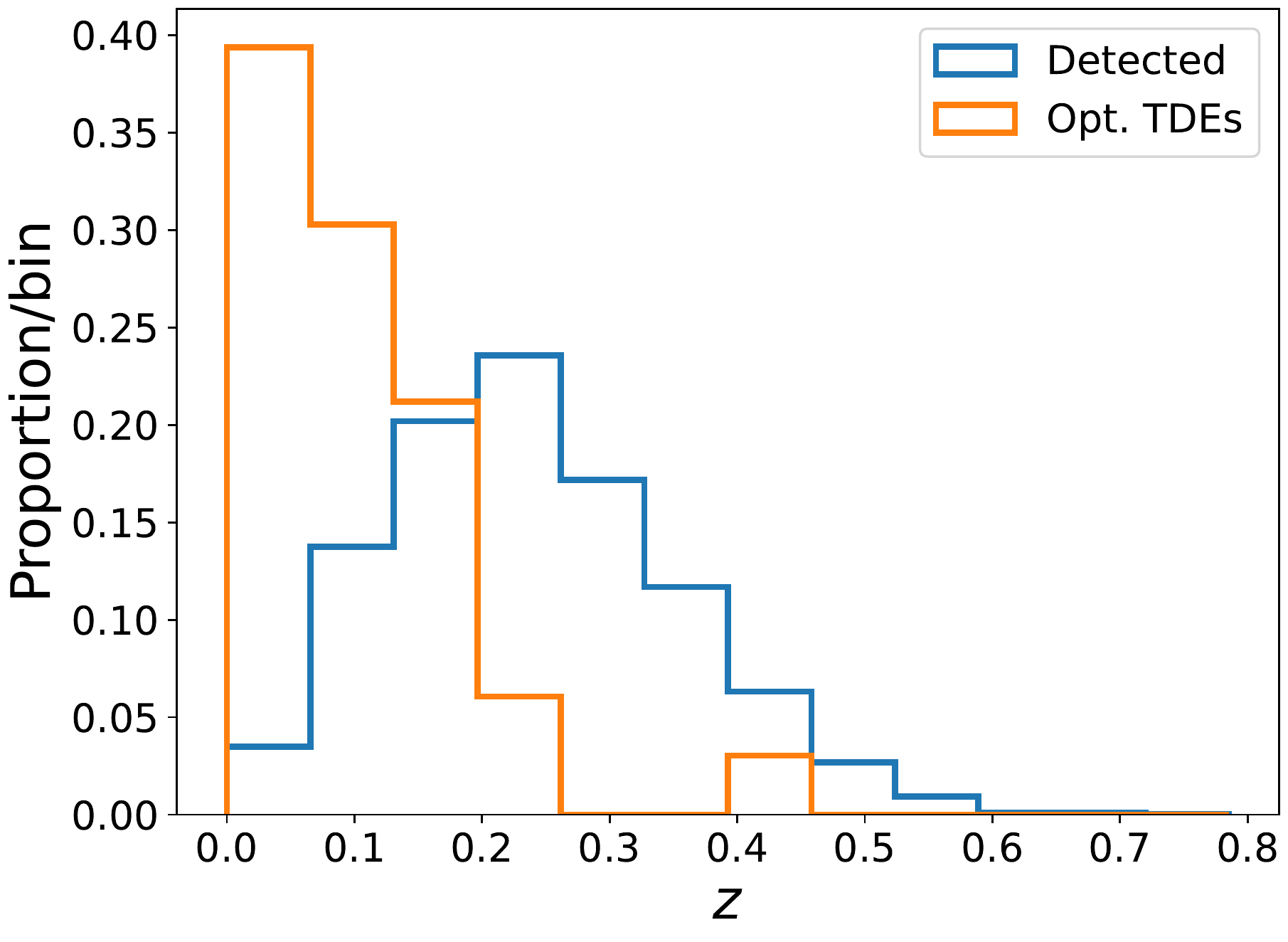}
    \caption{The histogram of TDE redshift distribution according to our 10$\times$10$\times$10 mock observations of the 440 deg$^2$ sky region. The blue histogram shows the 
    result from the mock detection based on 100-second reference images, while the orange histogram represents the distribution of the 33 optical-selected TDEs reviewed in Table~1 of \citet{2020SSRv..216..124V}. Compared with real sample, WFST shows great performance in detecting TDEs at higher redshifts, up to $z\sim0.8$.}
    \label{fig:zdist}
\end{figure}

The redshift distribution of TDEs detected in our mock observations is shown in Figure \ref{fig:zdist}. The higher-redshift TDEs call for deeper imaging, 
not only for the detection of the TDF emission itself, but also for the host galaxies, which is essential for identifying TDE candidates in the real observations. If the host galaxy detection is performed along with 
the 100-second exposure image, the redshift of TDEs can be up to $z\sim0.8$ (median $\approx0.23$), that is already higher than the most distant optical TDE, AT2020riz at $z=0.435$, found to date. 

\begin{figure*}
    \centering
    \includegraphics[width=0.7\textwidth]{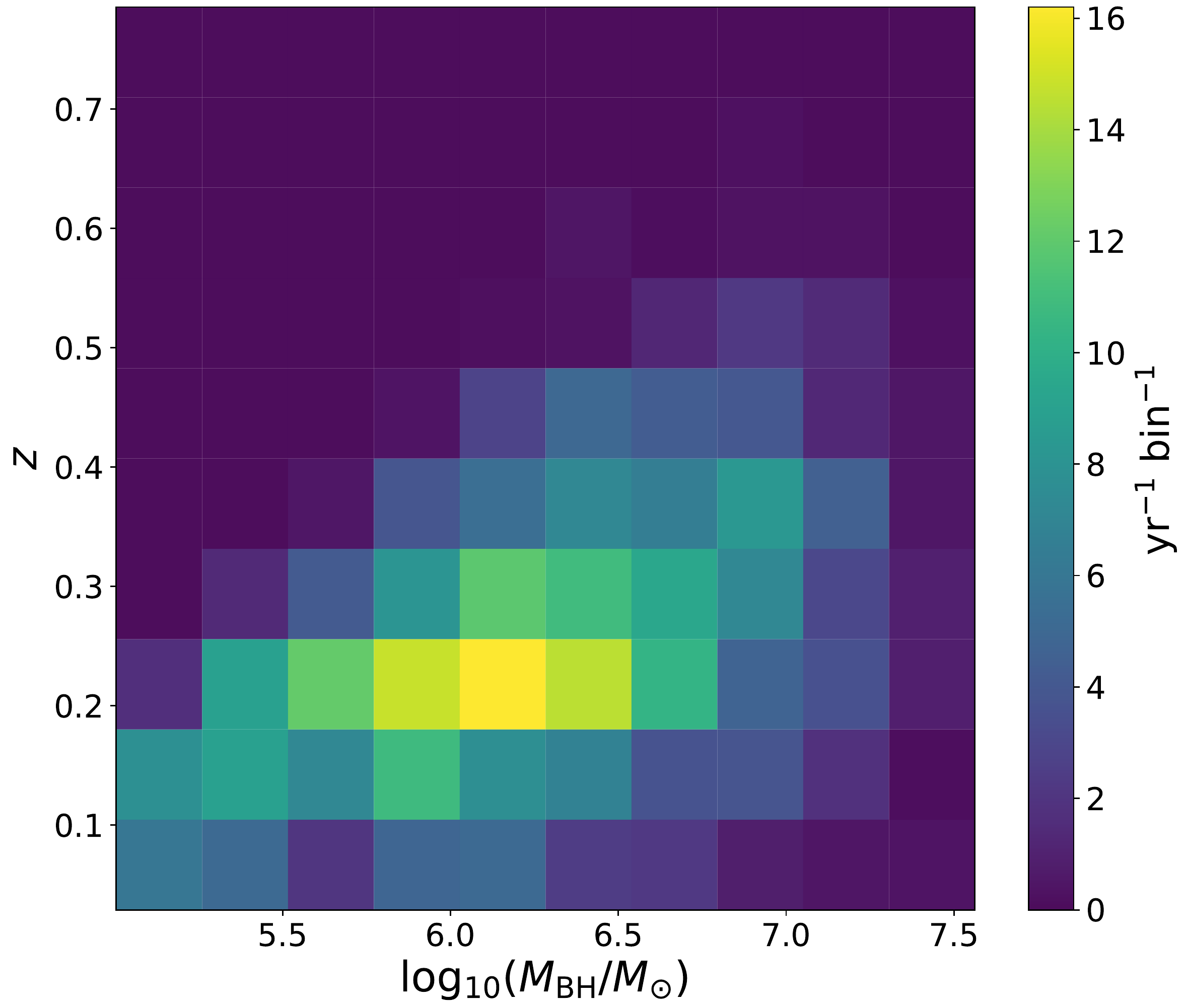}
    \caption{The $z$ and $\mbh$ distribution of detected TDEs. To minimize the stochastic errors, we show the average count of detectable TDEs in the 10$\times$10$\times$10 mock observations, that is equivalent to a one-year observation in sky area of 440 deg$^2$. TDEs occurring in smaller BHs, which usually resides in more dwarf galaxies, can be only detected at lower redshift.} 
    \label{fig:zmbhn}
\end{figure*}

We show the joint distribution of $z$ and $\mbh$ in Figure~\ref{fig:zmbhn}.
The detected TDEs are mainly distributed in the region of $z\sim0.2-0.3$ and $\mbh\sim10^6-10^7\,\msun$, that is roughly consistent with the prediction of VRO/LSST observations (\citealt{2021ApJ...910...93R}, their Figure~24). Obviously, $\mbh$ tends to be lower at lower redshifts. It can be readily understood
as a natural result of BH-host galaxy correlation since low-mass BHs usually reside in dwarf galaxies. More fainter galaxies are more easily detected in lower redshifts.

\subsubsection{$M_g$ distribution}

\begin{figure*}
    \centering
    \includegraphics[width=0.9\textwidth]{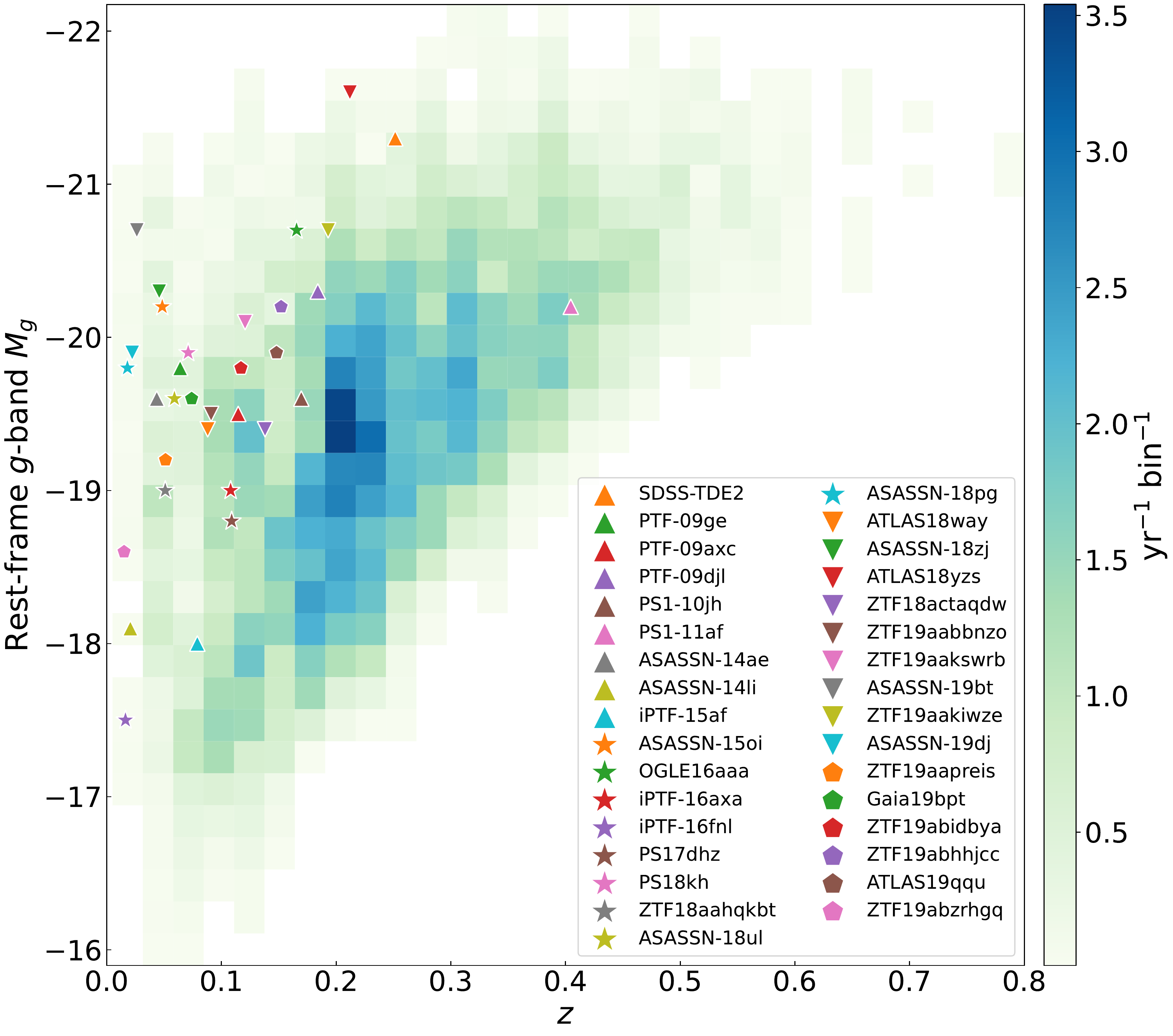}
    \caption{The peak absolute $g$-band magnitude ($M_g$) as function of redshift for TDEs detected in our mock observations. To minimize the stochastic errors, we show the average count of detectable TDEs in the 10$\times$10$\times$10 mock observations, that is equivalent to a one-year observation in sky area of 440 deg$^2$.
    We have also overplotted the 33 optical TDEs summarized in Table~1 of \citet{2020SSRv..216..124V} for comparison. The distribution of mocked TDE
    sample is generally in accordance with real optical TDEs, especially in luminosity, yet it has extended to higher redshift (see also Figure~\ref{fig:zdist}).}
    \label{fig:mg}
\end{figure*}

In Figure \ref{fig:mg} we show the distribution of absolute $g$-band magnitude ($M_g$) 
as a function of redshift. The flux limit determines that only brighter TDEs can be 
detected at higher redshift. Our TDE sample has kept its maximum luminosity around that of the brightest optical TDE, ATLAS18yzs, indicating reasonable parameter settings. Besides, it has extended to higher redshift than known optical TDEs.

The sparser distribution of our sample at $z<0.1$ is probably due to the small number 
of galaxies at such low redshift. Actually, galaxies at $z<0.1$ only occupy $\sim0.07\%$ 
among the total 52 million ones, leading to the very few TDEs found in a small
sky region (440 $\rm \deg^2$).

\subsubsection{Luminosity function}

\begin{figure*}
    \centering
    \includegraphics[width=0.9\textwidth]{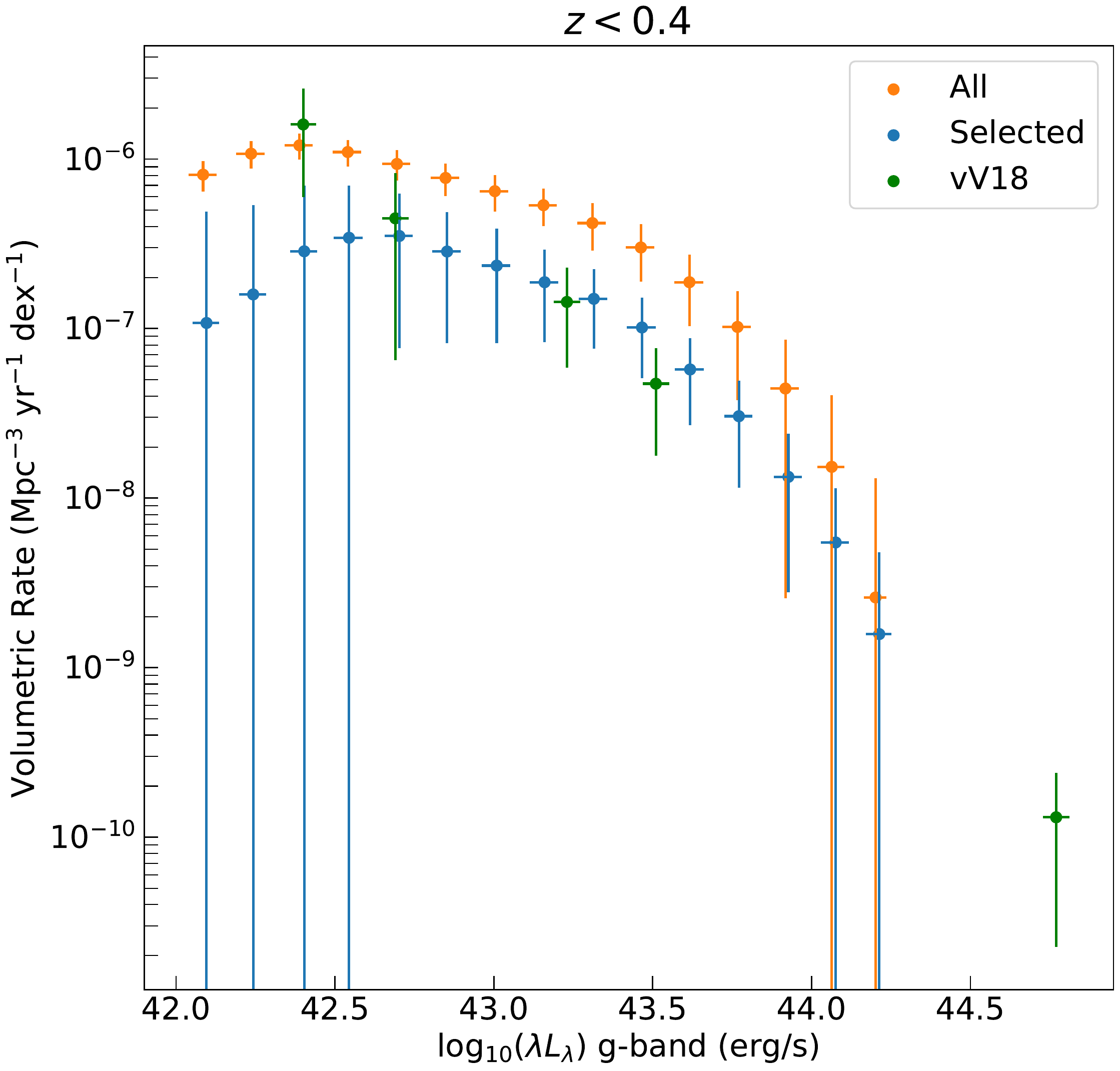}
    \caption{The luminosity function of detected TDFs (blue) and all TDFs (orange) within $z=0.4$ for 1000 mock observations. For detected mock TDFs, we perform the volumetric correction, i.e., "1/$V_{\mrm{max}}$" method following \citet{2018ApJ...852...72V}. The luminosity function of \citet{2018ApJ...852...72V} is also overplotted in green for comparison.}
    \label{fig:lf}
\end{figure*}

To further examine the reliability of the results, we have calculated the luminosity function (LF) of the detected TDEs at $z<0.4$ using the "1/$V_{\mrm{max}}$" method introduced in \citet{2018ApJ...852...72V},
and find it is roughly in agreement with the results of \citet{2018ApJ...852...72V}, as shown in Figure \ref{fig:lf}.
In the future WFST surveys, we may obtain a better real LF
basing on larger TDE samples, which will certainly promote our understanding 
of the optical emission of TDEs.

\subsubsection{Notes on the predicted detection rate}
If we choose an observational cadence of 10 days with $u,\ g,\ r,\ i$ filters, WFST can find 29 $\pm$ 6 TDEs per year in 440 deg$^2$ under the sky background level of 22 mag/arcsec$^2$. To examine the reliability of the predicted rate, we use the scaling relation from \citet{2011ApJ...741...73V}, that the rate should obey $\dot{N}_{\rm obs}\propto f_{\rm sky}F_{\rm lim}^{-3/2}$, where $f_{\rm sky}$ and $F_{\rm lim}$ represent the survey fraction of the whole sky and flux limit, respectively. As a reference, ZTF-I survey has discovered 30 TDEs in 15,000 deg$^2$ sky area in 32 months, indicating a discovery rate of $\sim$12 yr$^{-1}$ \citep{2022arXiv220301461H}. Assuming the flux limit of ZTF and WFST are both determined by $r$-band limiting magnitude, which is 20.5 mag and 22.8 mag, respectively, WFST should discover $\sim$8 TDEs per year in the 440 deg$^2$ field. This is a factor of $\sim$3.5 below our prediction. As ZTF-I TDEs all have follow-up spectroscopic, X-ray and UV data, it may suggest that some TDEs have been prevented from final identification due to the limited follow-up resources.

We turn to “the top of the iceberg,” the ZTF Bright Transient Survey \citep{2020ApJ...904...35P}, which covers the same sky area as ZTF. It is spectroscopically complete down to 18.5 mag, and has detected 5 TDEs in 25.5 months. We examine the scaling relation again, and it yields a detection rate of $\sim$26 TDEs per year in the 440 deg$^2$ field. This is now consistent with our prediction.

In addition, the theoretical TDE rate formula we use (Eq. \ref{eqn:dntde}) has probably contributed to an overestimate of the TDE detection rate. The reason is that the formula accounts for all TDEs. We use the MOSFiT model to generate light curves for only optically-bright TDEs. Therefore, the TDE rate assigned is precise only when all TDEs are optically-bright. However, a recent X-ray selected sample from eROSITA has unveiled a new class of TDEs, which are X-ray-bright but optically-faint \citep{2021MNRAS.508.3820S}. Despite this omission, the TDE rate estimated from eROSITA sample is $(1.1\pm0.5)\times10^{-5}$ gal$^{-1}$ yr$^{-1}$, that is one order of magnitude lower than $\sim$10$^{-4}$ gal$^{-1}$ yr$^{-1}$, as suggested by both optical TDE sample \citep{2020SSRv..216..124V} and the formula we use (both the results of \citet{2016MNRAS.455..859S} and Section \ref{sec:select}). Besides, the connections between X-ray and optical emission of TDEs remain unclear, as some TDEs are both detectable in X-ray and optical bands. Actually, among 13 eROSITA TDEs, 4 are detectable in optical surveys, and thus can be categorized into optically-bright TDEs. Therefore, we conclude that the X-ray-bright and optically-faint population occupy only a small fraction (<10\%) of the whole TDE family and thus they will not affect our results significantly.

\begin{figure*}
    \centering
    \includegraphics[width=0.75\linewidth]{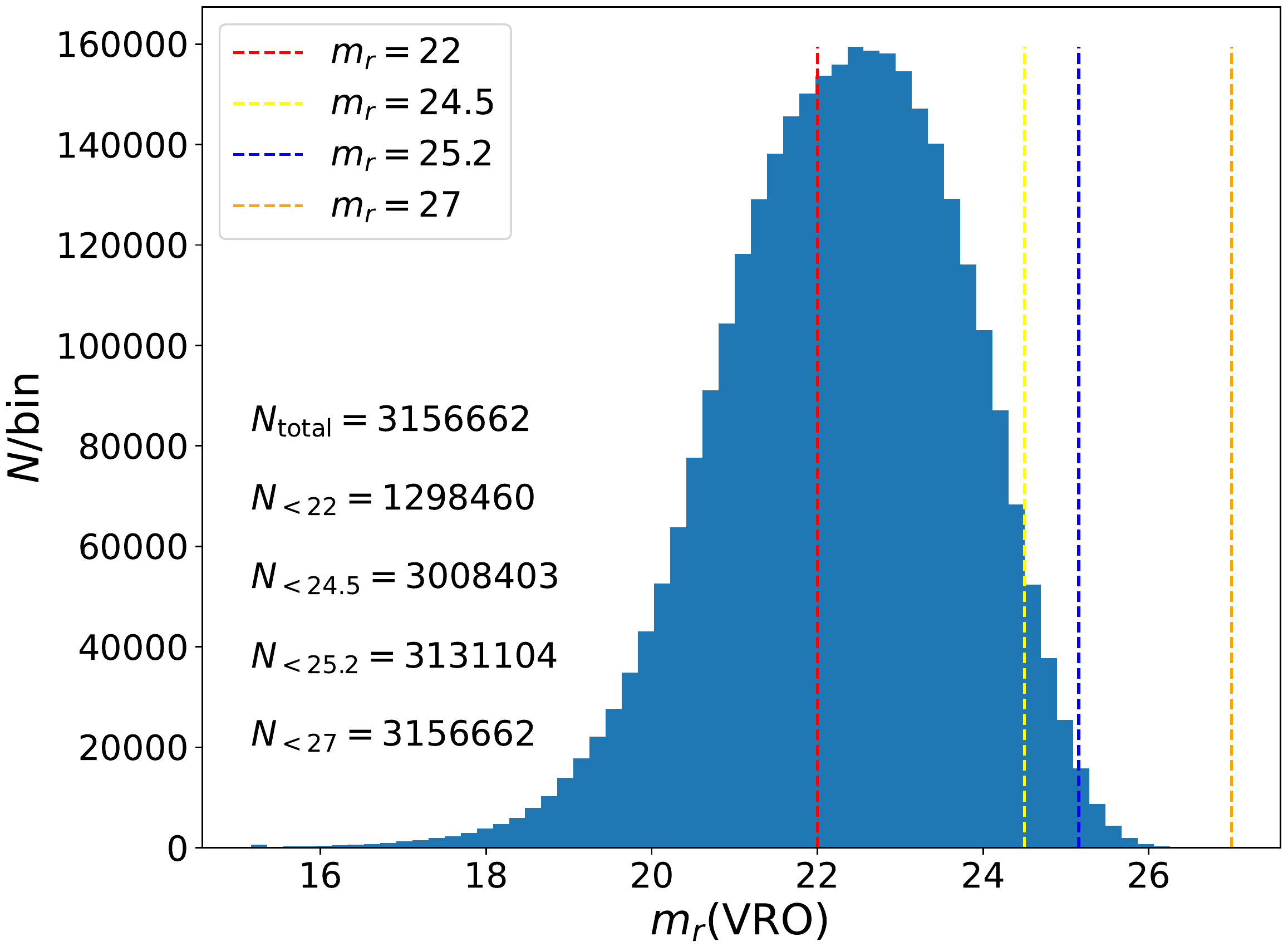}
    \caption{The $m_r$(VRO) distribution for galaxies with $10^5\,\msun\leqslant\mbh\leqslant9\times10^7\,\msun$ and $z<0.45$ in the CosmoDC2 catalogue, that matches the black hole mass and redshift ranges available for the TDE hosts, as presented in the Figure 24 of \citet{2021ApJ...910...93R}. The flux-limited catalogue they used excludes any flares coming from galaxies with $m_r>22$. Given the VRO depth ($m_r=24.5$ and 25.2 under 30-second and 100-second exposures, respectively, and $m_r\approx27$ after the 10-year survey) and $z<0.45$, the number of galaxies will be $\approx$1.5 times more by choosing the magnitude cut of $m_r=24.5$ instead of $m_r=22$. Moreover, the redshift limit $z_{\mrm{lim}}=0.45$ can actually be higher given a greater limiting magnitude, resulting in a more serious omission of possible TDE hosts ($\approx$4.5 times if $z_{\mrm{lim}}=0.8$). And if the host galaxy detection is performed on stacked images, the number of detectable TDEs will be even larger. Therefore, the predicted VRO detection rate given in \citet{2021ApJ...910...93R} is severely underestimated because they have put a very shallow magnitude cut for host galaxies, although the cut is reasonable for ZTF ($\approx$30\% for $z_{\mrm{lim}}=0.25$).}
    \label{fig:mr}
\end{figure*}

Given the WFST scanning speed (assuming 4-hour useful observing time per night and 10-second readout time), we can calculate the maximum survey area for this survey strategy. In this way, we estimate that $392\pm74$ TDEs per year can be discovered if the survey program is fully optimized for TDE search. This rate is about an order less than VRO/LSST predictions. For example, \citet{2021ApJ...910...93R} recently gives a very conservative detection rate of $\sim$1,600/yr for VRO/LSST. However, they have abandoned all flares in galaxies with $m_r>22$, which is far shallower than the depth of VRO. As a result, they have missed most TDEs in dwarf or distant galaxies (see Figure~\ref{fig:mr}) and thus the real detection ability of VRO could be likely even more powerful. In any case, in the upcoming VRO and WFST era, the anticipated TDE candidates will be massive and challenge the follow-up resources, such as spectroscopic observations.

\section{Conclusions} \label{sec:disc}


The 2.5-metre WFST is going to start a wide-field, fast and deep time-domain survey soon. In order to evaluate its survey ability, we have carried out mock observations of TDEs, and explored the influence of various factors. Our main results are summarized as below.

\begin{enumerate}

\item[$\bullet$] We define a discovery of TDE as $\geq$10 epochal detections in at least two bands. If we choose an observational cadence of 10 days with $u,\ g,\ r,\ i$ filters, WFST can find 29 $\pm$ 6 TDEs per year in 440 deg$^2$ under the sky background level of 22 mag/arcsec$^2$. Given the WFST scanning speed (assuming 4-hour useful observing time per night and 10-second readout time), we can calculate the maximum survey area for this survey strategy. In this way, we estimate that $392\pm74$ TDEs per year can be discovered if the survey program is fully optimized for TDE search.

\item[$\bullet$] $g$ and $r$ bands are most useful for the TDE detection; $u$ band cannot significantly increase the detection rate when both $g$ and $r$ bands are already used. However, it should still be important since it provides the radiation information closest to the SED peak of TDEs, which is crucial for distinguishing TDEs from other transients (e.g., by its blackbody temperature).

\item[$\bullet$] The sky background affects the results only when $I_{\mrm{sky}}<20$ mag/arcsec$^2$. The readout noise slightly prevents high-redshift TDEs from detection. 

\item[$\bullet$] Our mock observation yields TDEs up to $z\sim0.8$, with the largest number density at $z\sim0.2-0.3$. The $g$-band absolute magnitude ($M_g$) distribution and the luminosity function roughly agree with the real optical TDE sample.

\end{enumerate}


\section*{Acknowledgements}

We thank the expert referee for his/her very constructive comments and suggestions. We thank Tinggui Wang, Lulu Fan, Zhengyan Liu, Zesen Lin and Zhixiong Liang for very useful discussions and suggestions. This work is supported by the Strategic Priority Research Program of Chinese Academy of Sciences (No. XDB 41000000), the National Key R\&D Program of China (2017YFA0402600), the NSFC grant (Nos. 11833007, 11973038, 12073025), the Chinese Space Station Telescope (CSST) Project and the Fundamental Research Funds for the Central Universities. 

\section*{Data Availability}
The data underlying this article will be shared on reasonable request to the corresponding author. 







\appendix

\section{Detailed version of Table~\ref{TAB:THRCAD}}
\begin{table*}
\begin{threeparttable}
\caption{Full results of 10$\times$10$\times$10 mock observations\label{tab:full}}
\begin{tabular}{c|c|cccccccc}
\hline
\multicolumn{2}{c|}{Threshold $k$}  & 10  & 10  & 5   & 5   \\\hline
\multicolumn{2}{c|}{Cadence $C$(d)} & 10  & 20  & 10  & 20  \\\hline
\multicolumn{2}{c|}{$l_2$}          & 1   & 1   & 1   & 1   \\\hline
& $ug$ & 10.8$\pm$3.3(289.8$\pm$89.5) & 3.9$\pm$2.1(182.9$\pm$99.4) & 16.0$\pm$4.0(427.8$\pm$106.4) & 11.0$\pm$3.3(517.3$\pm$156.7) \\
& $ur$ & 17.2$\pm$4.3(459.9$\pm$115.5) & 5.0$\pm$2.4(232.7$\pm$112.7) & 27.3$\pm$5.5(732.0$\pm$148.5) & 17.8$\pm$4.4(832.7$\pm$205.5) \\
& $ui$ & 13.9$\pm$3.9(373.2$\pm$105.7) & 3.7$\pm$2.1(172.0$\pm$96.5) & 22.8$\pm$5.0(611.6$\pm$134.7) & 14.4$\pm$3.9(673.9$\pm$183.6) \\
& $gr$ & 28.3$\pm$5.4(757.1$\pm$145.4) & 10.2$\pm$3.4(479.3$\pm$159.7) & 41.3$\pm$6.3(1107.9$\pm$168.9) & 29.1$\pm$5.5(1362.5$\pm$256.5) \\
& $gi$ & 16.6$\pm$4.2(445.3$\pm$111.5) & 4.7$\pm$2.2(219.1$\pm$104.6) & 26.6$\pm$5.1(712.8$\pm$136.2) & 17.2$\pm$4.1(806.6$\pm$194.3) \\
Used & $ri$ & 17.4$\pm$4.0(466.4$\pm$108.3) & 4.8$\pm$2.3(223.7$\pm$107.0) & 28.1$\pm$5.1(753.9$\pm$135.6) & 18.1$\pm$4.0(846.2$\pm$188.0) \\
Bands& $ugr$ & 28.7$\pm$5.4(512.2$\pm$97.3) & 10.6$\pm$3.5(378.1$\pm$126.0) & 41.8$\pm$6.3(747.3$\pm$113.4) & 29.5$\pm$5.5(1054.1$\pm$196.4) \\
& $ugi$ & 19.0$\pm$4.6(339.5$\pm$81.4) & 6.0$\pm$2.6(213.4$\pm$92.1) & 29.5$\pm$5.4(527.6$\pm$97.0) & 19.6$\pm$4.6(702.0$\pm$164.7) \\
& $uri$ & 20.0$\pm$4.5(358.0$\pm$80.2) & 6.0$\pm$2.5(212.6$\pm$91.0) & 31.5$\pm$5.6(563.3$\pm$99.9) & 20.8$\pm$4.6(743.4$\pm$162.7) \\
& $gri$ & 28.8$\pm$5.5(515.2$\pm$97.8) & 10.3$\pm$3.4(368.8$\pm$122.1) & 42.3$\pm$6.2(755.5$\pm$111.2) & 29.7$\pm$5.5(1060.5$\pm$196.1) \\
& $ugri$ & 29.3$\pm$5.5(392.0$\pm$73.5) & 10.7$\pm$3.5(286.2$\pm$94.8) & 42.8$\pm$6.3(573.2$\pm$83.9) & 30.1$\pm$5.5(807.1$\pm$147.6) \\
& $ugriz$ & 29.3$\pm$5.5(313.6$\pm$58.8) & 10.7$\pm$3.5(229.0$\pm$75.8) & 42.8$\pm$6.3(458.6$\pm$67.1) & 30.1$\pm$5.5(645.7$\pm$118.1) \\
\hline
\hline
\multicolumn{6}{c}{}\\
\multicolumn{6}{c}{}\\
\hline
\multicolumn{2}{c|}{Threshold $k$}  & 10  & 10  & 5   & 5  \\ \hline
\multicolumn{2}{c|}{Cadence $C$(d)} & 10  & 20  & 10  & 20 \\ \hline
\multicolumn{2}{c|}{$l_2$}          & 2   & 2   & 2   & 2  \\ \hline
& $ug$ & 1.2$\pm$1.1(32.4$\pm$28.4) & 0.5$\pm$0.6(22.9$\pm$29.3) & 1.7$\pm$1.3(46.1$\pm$34.7) & 1.2$\pm$1.1(57.9$\pm$49.9)\\
& $ur$ & 1.0$\pm$0.9(25.5$\pm$25.2) & 0.3$\pm$0.5(15.1$\pm$24.7) & 1.4$\pm$1.2(38.7$\pm$31.7) & 1.0$\pm$0.9(45.4$\pm$43.7)\\
& $ui$ & 0.8$\pm$0.8(20.7$\pm$22.4) & 0.2$\pm$0.5(10.1$\pm$21.2) & 1.2$\pm$1.1(32.6$\pm$28.9) & 0.8$\pm$0.8(37.5$\pm$39.5)\\
& $gr$ & 14.9$\pm$4.1(399.9$\pm$108.6) & 6.2$\pm$2.7(292.3$\pm$125.9) & 20.8$\pm$4.7(557.9$\pm$124.7) & 15.3$\pm$4.1(715.3$\pm$192.9)\\
& $gi$ & 11.0$\pm$3.3(294.5$\pm$87.4) & 3.6$\pm$2.0(168.5$\pm$91.5) & 16.6$\pm$4.0(445.6$\pm$106.1) & 11.3$\pm$3.3(529.6$\pm$155.2)\\
Used & $ri$ & 16.5$\pm$4.1(443.1$\pm$111.0) & 4.6$\pm$2.2(217.8$\pm$104.2) & 26.3$\pm$5.1(704.9$\pm$135.8) & 17.1$\pm$4.1(801.0$\pm$193.8)\\
Bands& $ugr$ & 15.3$\pm$4.1(274.0$\pm$73.1) & 6.6$\pm$2.8(235.5$\pm$99.9) & 21.3$\pm$4.6(380.6$\pm$82.9) & 15.7$\pm$4.1(560.8$\pm$147.8)\\
& $ugi$ & 12.4$\pm$3.6(221.3$\pm$63.9) & 4.5$\pm$2.2(162.4$\pm$79.1) & 18.1$\pm$4.2(323.0$\pm$74.4) & 12.7$\pm$3.6(453.3$\pm$128.0)\\
& $uri$ & 18.4$\pm$4.5(328.4$\pm$80.3) & 5.6$\pm$2.4(199.1$\pm$87.4) & 28.6$\pm$5.4(510.7$\pm$96.7) & 19.0$\pm$4.6(677.4$\pm$163.2)\\
& $gri$ & 25.4$\pm$5.2(454.6$\pm$92.5) & 9.4$\pm$3.4(334.4$\pm$120.7) & 37.0$\pm$5.7(661.4$\pm$101.5) & 26.2$\pm$5.2(935.6$\pm$185.8)\\
& $ugri$ & 25.9$\pm$5.2(346.5$\pm$70.0) & 9.7$\pm$3.5(260.4$\pm$93.8) & 37.5$\pm$5.7(502.6$\pm$76.6) & 26.6$\pm$5.2(713.5$\pm$140.6)\\
& $ugriz$ & 25.9$\pm$5.2(277.2$\pm$56.0) & 9.7$\pm$3.5(208.3$\pm$75.0) & 37.5$\pm$5.7(402.1$\pm$61.3) & 26.6$\pm$5.2(570.8$\pm$112.5)\\
\hline
\end{tabular}
\begin{tablenotes}
\item \textbf{Notes:}
\item 1. Out of parentheses: \\Detectable TDEs in 440 deg$^{2}$ per year, with threshold, cadence, the minimum number of WFST bands that can detect the host galaxy and band combination altered.
\item 2. In parentheses: \\Detectable TDEs in maximum survey area per year, with threshold, cadence, the minimum number of WFST bands that can detect the host galaxy and band combination altered. The maximum survey area are calculated based on the WFST scanning speed (assuming 4-hour useful observing time per night, 10-second readout time), with an upper limit of 2$\pi$ steradian ($\sim$20,600 deg$^2$) if the band combination can fully cover the full northern celestial sphere.
\item 3. Explanations on key parameters:\\
Threshold $k$: The minimum times that the TDF should be detected in a band of WFST. \\
Cadence $C$: The interval between two closest observations in any band of WFST.\\
$l_2$: The minimum number of WFST bands that can detect the host galaxy.
\end{tablenotes}
\end{threeparttable}
\end{table*}


\bsp	
\label{lastpage}
\end{document}